\documentclass[aps,prb,floats,showpacs,preprintnumbers,10pt]{revtex4}
\usepackage{natbib}
\usepackage{amsmath}
\usepackage{amsfonts}
\usepackage{amssymb}
\usepackage{graphicx}
\usepackage{dcolumn}
\usepackage{bm}
\usepackage{subfigure}
\usepackage{textcomp}
\usepackage{enumerate}

\begin{document}
\title{\textit{Simulating the radiofrequency dielectric response of relaxor ferroelectrics: Combination of Coarse-Grained Hamiltonians and Kinetic Monte Carlo}}

\author{Gr\'egory Geneste$^1$}
\email{gregory.geneste@cea.fr}
\author{L. Bellaiche$^2$}
\author{Jean-Michel Kiat$^{3,4}$}

\affiliation{$^1$ CEA, DAM, DIF, F-91297 Arpajon, France}
\affiliation{$^2$ Physics Department and Institute for Nanoscience and Engineering, University of Arkansas, Fayetteville, Arkansas 72701, USA}
\affiliation{$^3$ Laboratoire Structures, Propri\'et\'es et Mod\'elisation des Solides, Universit\'e Paris Saclay, CentraleSup\'elec, CNRS (UMR 8580), Grande voie des vignes, 92295 Ch\^atenay-Malabry}
\affiliation{$^4$ LLB, CEA, CNRS, Universit\'e Paris-Saclay 91191 Gif-sur-Yvette, France}

\date{\today}


\begin{abstract}
The radiofrequency dielectric response of the lead-free Ba(Zr$_{0.5}$Ti$_{0.5}$)O$_3$ relaxor ferroelectric is simulated using a coarse-grained Hamiltonian. This concept, taken from Real-Space Renormalization Group theories, allows depicting the collective behavior of correlated local modes gathered in blocks. Free-energy barriers for their thermally activated collective hopping are deduced from this {\it ab-initio}-based approach, and used as input data of Kinetic Monte Carlo simulations. The resulting numerical scheme allows to simulate the dielectric response for external field frequencies ranging from the kHz up to a few tens of MHz for the first time, and to, e.g., demonstrate that local (electric or elastic) random fields lead to the dielectric relaxation in the radiofrequency range that has been observed in relaxors.
\end{abstract}

\maketitle

Relaxors with perovskite structure form an important family of functional materials that exhibit intriguing dielectric properties~\cite{burns1983,cross1987}: the real part of the frequency-dependent dielectric permittivity has a maximum with temperature, at $T_{max}$, while the system remains macroscopically paraelectric down to the lowest temperature, and $T_{max}$ depends on the frequency of the applied electric field, a phenomenon called {\it dielectric relaxation}. Different suggestions have been proposed to explain these macroscopic properties, such as non-local (electric or elastic) random fields (RFs)~\cite{kleemann1} (electric or elastic fields {\it on site $i$} depending on the chemical disorder {\it surrounding $i$}), and the possible existence, and interplay, of polar nanoregions (PNRs), i.e. polar instabilities that correlate the elementary dipoles on a few lattice constants. The location and properties of these PNRs would be dependent on the local chemical disorder, that relaxors can exhibit on one of their sublattices~\cite{jmk}.

The dynamics of the electric dipoles of such structures is believed to be associated with characteristic times being much larger than typical atomic times, and being temperature-dependent (as a result of thermal activation). These large time scales are responsible for the frequency-dependence of the dielectric permittivity in the radiofrequency domain (from the kHz up to several tens of MHz). Recently, microscopic description of relaxors, based on model Hamiltonians derived from first-principles coupled to Monte Carlo (MC) or Molecular Dynamics (MD) simulations, have provided precious information about the effect of RFs on relaxor properties and the nature of these PNRs~\cite{burton1,burton2,burton3,burton4,bzt,bzt2,grinberg2007,grinberg2009}. In heterovalent relaxors such as PbMg$_{1/3}$Nb$_{2/3}$O$_3$ (PMN)~\cite{pmn1,pmn2,pmn3,pmn4}, the PNRs are suggested to arise from complex phenomena including strong non-local electric RFs~\cite{pmn,phelan2014}. By contrast, in homovalent relaxors such as Ba(Zr,Ti)O$_3$ (BZT), Ref.~\cite{bzt} numerically found that PNRs appear in regions where the chemical species driving the polar instability (Ti) is more abundant, i.e., it is the {\it local} RFs arising from the difference in polarizability between Ti and Zr ions that induce relaxor behavior, while non-local electric and elastic RFs have a rather negligible effect. Note that local RFs can lead to very long relaxation times in disordered magnets~\cite{fisher1986}, which may also be the case for relaxors~\cite{pmn3}.

In order to gain a further deeper understanding of relaxor ferroelectrics, it is highly desired to have numerical schemes able to simulate the most striking characteristics of relaxors, i.e. the radiofrequency dielectric relaxation. However, to the best of our knowledge, such schemes do not exist. One reason behind this paucity is that MD simulations are limited to a few nanoseconds, and thus cannot give access to the time scales required to mimic the radiofrequency dielectric response of relaxors. However, the Kinetic Monte Carlo (KMC) method, that we recently applied to simulate the radiofrequency dielectric response of Li-doped KTaO$_3$ (KLT)~\cite{kmcktl2011}, is able to reproduce such time scales.
Nevertheless, in KLT, the elementary processes driving the dielectric response involve few degrees of freedom (hoppings of individual Li impurities), with rather temperature-independent energy barriers~\cite{kmcktl2011}, two assumptions clearly not obeyed in relaxor ferroelectrics as, e.g. evidenced by the fact that PNRs do not exist anymore above the Burns temperature, and that the processes responsible for the dielectric response involve the {\it collective motion} of several microscopic degrees of freedom, since a PNR should extend over several unit cells.

In this Letter, we report the development, and results, of a novel numerical approach able to simulate the radiofrequency dielectric response of relaxors. It is based on a Renormalization Group (RG) transformation in real space~\cite{introcritical}, combined with KMC. In particular, this new scheme allows describing radiofrequency dielectric relaxation in BZT compatible with the Vogel-F\"ulcher law~\cite{vf1,vf2}, therefore demonstrating the power and promise of such methodology, and that non-local electric or elastic RFs are not absolute requirements to generate relaxor behaviors. In other words, local RFs are enough to induce relaxor properties.

\begin{figure}[htbp]
    {\par\centering
    {\scalebox{0.45}{\includegraphics{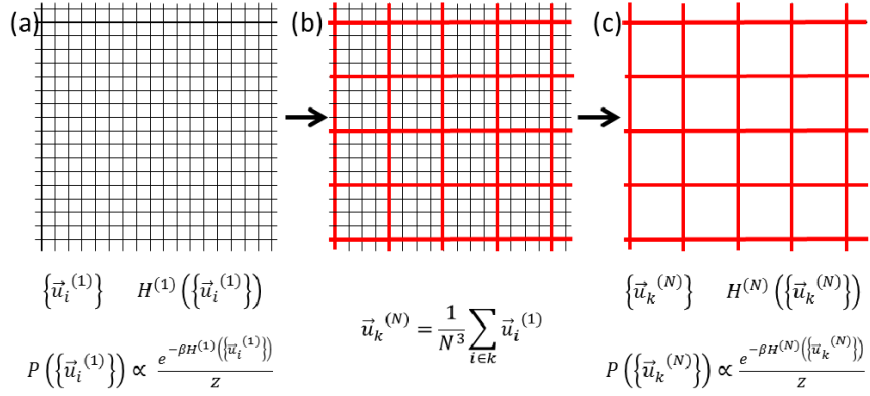}}}
    \par}
     \caption{{\small Coarse-graining process. (a): real system; the microscopic local modes interact through the microscopic Hamiltonian $H^{(1)}$; (b): the unit cells are gathered by blocks, and a block variable is defined in each block, as the mean local mode; (c): the coarse-grained Hamiltonian $H^{(N)}$ is constructed, preserving the partition function and the macroscopic observables. In case a (resp. c), $P$ denotes the density of probability of the microscopic state $\{ {\bf{u_i}}^{(1)} \}$ (resp. constrained-block state $\{ {\bf{u_k}}^{(N)} \}$).}}
    \label{fig1}
\end{figure}

We denote by $\{ {\bf{u_i}}^{(1)} \}$ the set of the microscopic local modes, and
associate to each elementary unit cell $i$ a real number $x_i^{(1)}$ characterizing its composition (0 for Zr, 1 for Ti).
The $\{ x_i^{(1)} \}$ play the role of a set of parameters (not variables) of the microscopic Hamiltonian~\cite{zhong1995,zhong1994,afd,bfo}, for which we use the following form:

\begin{eqnarray}
\label{heff2}
H^{(1)}(\{ {\bf{u_i}}^{(1)} \};\{x_i^{(1)} \})=\underbrace{\sum_{i} H_{loc}^{(1)}({\bf{u_i}}^{(1)};x_i^{(1)})}_{local~random~fields}
+ \frac{1}{2} \sum_{i, j, \alpha, \beta \atop (i \ne j)}  \bar{\bar{C}}_{SR,\alpha \beta}^{(1)}(i,j) u_{i \alpha}^{(1)} u_{j \beta}^{(1)} 
+ \frac{1}{2} \sum_{i, j, \alpha, \beta \atop (i \ne j)}  \bar{\bar{C}}_{LR,\alpha \beta}^{(1)}(i,j) u_{i \alpha}^{(1)} u_{j \beta}^{(1)},
\end{eqnarray}

$\bar{\bar{C}}_{SR}^{(1)}(i,j)$ and $\bar{\bar{C}}_{LR}^{(1)}(i,j)$ being the matrices describing the short-range (SR) and dipole-dipole (LR) interaction between local modes in cells $i$ and $j$. 
In such form, the chemical disorder contributes only to the {\it local part}, 
characterizing {\it local} RFs arising from the polarizability difference between Ti and Zr ions. Refs.~\cite{bzt,wang2011} have shown that such RFs are sufficient to reproduce several properties of BZT such as the temperature evolution of the static and hyper frequency dielectric permittivity, and the existence of small PNRs -- which contrasts with heterovalent relaxors, for which {\it non-local} RFs (contributions of the
chemical disorder from neighboring cells and related to interaction terms of $H^{(1)}$), are large and play a fundamental role in the dielectric properties ~\cite{pmn}. 
We will show here, by neglecting non-local RFs, that local RFs in BZT can generate a radiofrequency dielectric relaxation by themselves.

Our approach consists in coarse-graining the system (Fig.~\ref{fig1}), i.e. dividing it in $N \times N \times N$ cubic blocks, and defining, in each block $k$, a "block variable" (or local order parameter) as the mean local mode over the block: ${\bf{u_k}}^{(N)} = \frac{1}{N^3} \sum_{i \in k} {\bf{u_i}}^{(1)}$. We also define a local composition in each block, as $x_k^{(N)} = \frac{1}{N^3} \sum_{i \in k} x_i^{(1)}$, which takes fractional values $\in$ [0;1], allowing to define Ti-rich blocks ($x \rightarrow$ 1) and Ti-poor ones ($x \rightarrow$ 0). The set of block variables $\{ {\bf{u_k}}^{(N)} \}$ is used to define an {\it incomplete partition function}~\cite{binder1987,geneste2009}, 

\begin{eqnarray}
\tilde{Z}^{(N)}(\{ {\bf{u_k}}^{(N)} \}; \{ x_i^{(1)} \})  = 
C \int ... \int \{ \prod_{blocks~k} \delta(\sum_{i \in k} {\bf{u_i}}^{(1)}-N^3 {\bf{u_k}}^{(N)}) \}
 e^{- \beta H^{(1)}(\{ {\bf{u_i}}^{(1)} \};\{ x_i^{(1)} \})} \{ \prod_{i} d {\bf{u_i}}^{(1)} \} ,
\end{eqnarray}

by summing over all the microscopic states $\{ {\bf{u_i}}^{(1)} \}$ such that $\forall k$, $\sum_{i \in k} {\bf{u_i}}^{(1)} = N^3 {\bf{u_k}}^{(N)}$. The $\{ {\bf{u_k}}^{(N)} \}$ are thus variables of $\tilde{Z}^{(N)}$.
The coarse-grained Hamiltonian~\cite{binder1987} $H^{(N)}$ is then defined, up to an additive constant, as 
$H^{(N)}(\{ {\bf{u_k}}^{(N)}\};\{ x_i^{(1)} \}) = - k_B T ln \tilde{Z}^{(N)}(\{ {\bf{u_k}}^{(N)} \};\{ x_i^{(1)} \} )$.
It has the physical meaning of an incomplete free energy, and depends on temperature (at contrast with $H^{(1)}$).

The system of the block variables interacting through $H^{(N)}$ has, by construction, the same partition function as the initial system (up to a multiplicative constant), and thus the same macroscopic observables. To calculate this coarse-grained Hamiltonian, we use the following formula~\cite{geneste2009,geneste2011,sprik1998}:

\begin{equation}
\label{derivative}
\frac{ \partial H^{(N)} }{\partial {\bf{u_n}}^{(N)}}  (\{ {\bf{u_k}}^{(N)} \};\{ x_i^{(1)} \}) = 
- < \sum_{i \in n} {\bf{f_i}} > (\{ {\bf{u_k}}^{(N)} \};\{ x_i^{(1)} \}),
\end{equation}

where $<...>(\{ {\bf{u_k}}^{(N)} \})$ denotes thermal conditional average~\cite{sprik1998} evaluated at fixed $\{ {\bf{u_k}}^{(N)} \}$ (see Suppl. Info).
${\bf{f_i}} = - \frac{\partial H^{(1)}}{\partial {\bf{u_i}}^{(1)}}$ is the force on local mode $i$. $H^{(N)}$ is a {\it potential of mean force}, as employed in chemical physics to study chemical reactions along a reaction coordinate~\cite{sprik1998, free-en1,free-en2,free-en3}. Here, the "reaction coordinate" is multidimensional and consists in the set of all the block variables, $\{ {\bf{u_k}}^{(N)} \}$.

This coarse-grained Hamiltonian can be approximated by a form identical to that of $H^{(1)}$, i.e.
$H^{(N)}  (\{ {\bf{u_k}}^{(N)} \}; \{ x_k^{(N)} \})$,
provided some hypothesis are fulfilled. The most important is that the block size $Na_0$ remains lower than the correlation length $\xi$(T), i.e. the local modes must be correlated together all over the block~\cite{troster2005,geneste2010}:

\begin{eqnarray}
\label{hn}
H^{(N)}  (\{ {\bf{u_k}}^{(N)} \}; \{ x_k^{(N)} \}) \approx
\sum_{n} H_{loc}^{(N)} ({\bf{u_n}}^{(N)}; x_n^{(N)} )   
+ \frac{1}{2} \sum_{n, n', \alpha, \beta \atop (n \ne n')} 
 \bar{\bar{C}}_{SR, \alpha \beta}^{(N)} (n,n') u_{n \alpha}^{(N)} u_{n' \beta}^{(N)}
+ \frac{1}{2} \sum_{n, n', \alpha, \beta \atop (n \ne n')} 
 \bar{\bar{C}}_{LR, \alpha \beta}^{(N)} (n,n') u_{n \alpha}^{(N)} u_{n' \beta}^{(N)}
\end{eqnarray}

The renormalized coefficients $\bar{\bar{C}}_{SR}^{(N)} (n,n')$ and $\bar{\bar{C}}_{LR}^{(N)} (n,n')$ describe SR and LR interactions between blocks $n$ and $n'$, while the local free energy $H_{loc}^{(N)} ({\bf{u}};x)$ describes locally the thermodynamics of a block having as local order parameter ${\bf u}$, and chemical composition $x$. The condition $Na_0 < \xi(T)$ ensures that the blocks contain at most one PNR. 
At given temperature, several values of $N$ are thus possible, but the most convenient is $Na_0 \sim \xi(T)$.
MC simulations using $H^{(1)}$ reveal that at low temperature, collective behavior of the local modes is observed within Ti-rich blocks of typical size~\cite{bzt} 2 $\times$ 2 $\times$ 2: $N$=2 is the block size used here. This may appear as small, but 
within Ti-rich blocks, collective motion of 6-8 local modes is sufficient to generate a radiofrequency dielectric response, as we will see.

\begin{figure}[htbp]
    {\par\centering
    {\scalebox{0.4}{\includegraphics{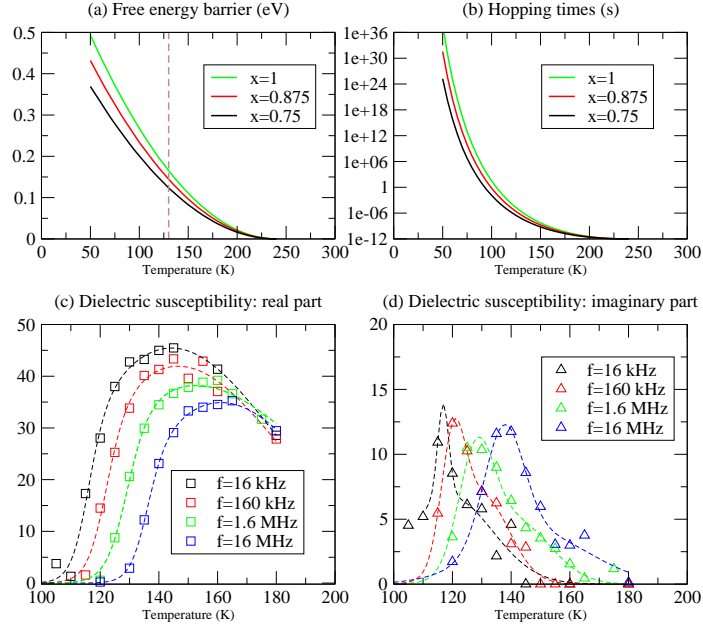}}}
   \par}
     \caption{{\small Upper panel: local free energy barriers $\Delta F_{loc}(T;x)$ (eV), and hopping times, $\tau(T;x) = \tau_0 e^{+ \Delta F_{loc}(T;x) / k_B T}$ (s), as a function of temperature for the three Ti-richest blocks ($x$=1, 0.875 and 0.75). Lower panel: temperature evolution of $\chi'$ and $\chi''$. In panel c, the lines correspond to a fit according to Ref.~\cite{cheng1998}. In panel d, the lines are guides for the eyes.}}
    \label{fig2}
\end{figure}

$\bar{\bar{C}}_{SR}^{(N)}$ and $\bar{\bar{C}}_{LR}^{(N)}$ are computed directly from the coefficients of $H^{(1)}$ (see Suppl. Info.). 
Ideally, the local free energy $H_{loc}^{(N)}$ could be obtained from constrained MD simulations~\cite{geneste2009,geneste2010,geneste2011,kumar2010}. Here, for simplicity, we use the following phenomenological form, expressed as an energy {\it per 5-atom cell}:
$H_{loc}^{(N)}({\bf{u}};x) - H_{loc}^{(N)}({\bf{0}};x) = x [ H_{loc,Ti}^{(N)}(\frac{{\bf{u}}}{x}) -  H_{loc,Ti}^{(N)}({\bf{0}}) ]$,
with $H_{loc,Ti}^{(N)}({\bf{u}}) - H_{loc,Ti}^{(N)}({\bf{0}}) = a_1' (T-T_0) (u_X^2 + u_Y^2 + u_Z^2) + a_{11} (u_X^4 + u_Y^4 + u_Z^4)$ the local free energy for a Ti-rich block ($x$=1). Within this form, all the blocks are polar under $T_0$, not above. The polar character increases with decreasing $T$, and with $x \rightarrow$ 1. The coefficients are extracted from microscopic MD simulations:
$T_0$ = 240 K (the so-called $T^*$ of BZT, according to Ref.~\onlinecite{bzt}), $a'_1$ = 0.053654 eV/(\AA$^2$.K), and $a_{11}$ = 420.8273 eV/\AA$^4$.
There are 8 local minima along the $<$111$>$ directions, and saddle points between a minimum and another are taken as the minima of $H_{loc}^{(N)}$ along $<$110$>$. The hopping local free energy barriers $\Delta F_{loc}(T;x)$ thus depend on temperature (Fig.~\ref{fig2}a) and on local chemical composition of the block. The associated transition rates are deduced from transition state theory~\cite{tst1,kramers}, as $r(T;x) = r_0 e^{- \Delta F_{loc}(T;x) / k_B T}$. The corresponding relaxation times $\tau(T;x) = \tau_0 e^{+ \Delta F_{loc}(T;x) / k_B T}$ are plotted in Fig.~\ref{fig2}b as a function of temperature. Above 240 K, static PNRs do not exist any more, while below $\sim$ 130 K (freezing temperature~\cite{bzt}), the relaxation times increase, tending rapidly but continuously towards a freezing of the polar blocks. The phenomenon of freezing, within this phenomenological description of $H_{loc}^{(N)}$, corresponds to a sharp, but continuous increase of the relaxation times (no divergence at $T_f$ for BZT).

Having the free energy landscape of the $\{ {\bf{u_k}}^{(N)} \}$, we can perform a KMC simulation of the dielectric response. A supercell of 12 $\times$ 12 $\times$ 12 blocks is constructed, starting from a 24 $\times$ 24 $\times$ 24 supercell of 5-atom cells in which the Ti and Zr are randomly distributed with equal probability. Each polar block $n$ has a local free energy surface depending on $T$ and chemical composition $x_n$. However, for physical and practical reasons, we do not consider as elementary events the hoppings in {\it all} the blocks, because Ti-rich blocks may have large free energy barriers, and the poorest ones, very small barriers. Thus, very rare events can coexist with very frequent ones, making the KMC algorithm untractable. Moreover, our phenomenological form for $H_{loc}^{(N)}$ does not stand for Zr-rich blocks, which remain non polar down to zero K~\cite{bzt}. The events kept for the KMC therefore correspond to the hopping times relevant to the radiofrequency dielectric response. This is equivalent to retain only the Ti-richest blocks as polar ($x$=1, 0.875 and 0.75), the other ones being considered as a dielectric medium. The system becomes thus equivalent to a dipole glass~\cite{sherrington} embedded in a dielectric matrix, as in KLT, at the exception that the energy barriers depend on temperature and on local chemical composition, and that each block variable has 8 possible local minima along $<$111$>$~\cite{bzt}.

\begin{figure}[htbp]
    {\par\centering
   {\scalebox{0.375}{\includegraphics{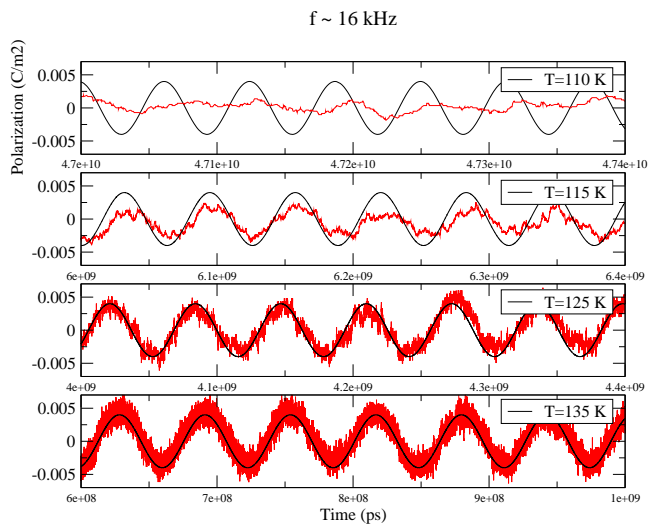}}}
   \par}
     \caption{{\small Example of time evolution of the macroscopic polarization (red curve, in C/m$^2$) under sinusoidal electric field (black curve, amplitude 1.0 $\times$ 10$^{+7}$ V/m) of frequency $\approx$ 16 kHz, for different temperatures.}}
    \label{fig3}
\end{figure}

The radiofrequency dielectric response is obtained by applying a sinusoidal external electric field along ${\bf{e_{x}}}$ (unit vector along the pseudo-cubic [100] axis): ${\bf{E_{ext}}}(t) = E_0 cos(\omega t) {\bf{e_{x}}}$. 
Each block variable $n$ feels a local field consisting of this external field plus an internal field ${\bf{E_{int}}} (n)$ associated with the (renormalized) SR and LR interactions between blocks: ${\bf{E_{loc}}}(n) = {\bf{E_{ext}}} (t) + {\bf{E_{int}}} (n)$. We assume that ${\bf{E_{loc}}}(n)$ does not modify the position of the transition states (saddle points of $H_{loc}^{(N)}$), but changes the free energy barrier from site $k$ to $k'$ according to $\Delta F_{loc}(T;x) \rightarrow  \Delta F_{loc}(T;x) + Z_{b}^{*} ({\bf{u_{k}}} - {\bf{u_{k,k'}}}). {\bf{E_{loc}}}(n)$, $Z_b^{*}$ being the block effective charge, ${\bf{u_{k}}}$ the position of stable site $k$ (among the 8 possible) and ${\bf{u_{k,k'}}}$ that of the transition state from $k$ to $k'$.

When ${\bf{E_{ext}}}$ is applied, the sites  ${\bf{u_{k}}}$ that are in the same direction as ${\bf{E_{ext}}}$ are stabilized, and the energy barriers towards these sites are lowered with respect to the barriers of the backwards motions. Thus the local order parameters tend to align along the external field, producing a macroscopic polarization $P_X$. An example at $f \approx$ 16 kHz is shown on Fig.~\ref{fig3}. However, this alignment occurs with a delay related to the relaxation time of the block variables, compared to the external field period ($t_e=2 \pi / \omega$), so that $P_X$ takes the form $P_X(t) = P_0(\omega) cos(\omega t + \phi(\omega))$ (assuming linear response). A fit is then performed, providing the amplitude $P_0(\omega)$ and the phase $\phi(\omega)$, from which the real and imaginary parts, $\chi'$ and $\chi''$, of the dielectric susceptibility are obtained, as
$\chi' = \frac{P_0(\omega)}{\epsilon_0 E_0} cos(\phi(\omega))$ and $\chi'' = \frac{P_0(\omega)}{\epsilon_0 E_0} | sin(\phi(\omega)) |$.

Figures~\ref{fig2}c and d show the temperature evolution of $\chi'$ and $\chi''$ in Ba(Zr$_{0.5}$Ti$_{0.5}$)O$_3$, for external field frequencies between 16 kHz and 16 MHz. As in KLT, the dielectric relaxation is well reproduced by the KMC, with a maximum of $\chi'$, $T_{max}(\omega)$ evolving with $\omega$, and the curves $\chi'(T)$ enveloping each other as $\omega$ decreases. The temperature $T_{max}'$ of the maximum of $\chi''$ are $< T_{max}$, as experimentally observed~\cite{dixit2006,kleemann2,usman2013}. The peak positions are in rather good agreement with experiments~\cite{nuzhnyy2012,maiti2008,usman2013,dixit2006}. The values of $\chi'$, however, are $\sim$ one order of magnitude too low compared to experiments~\cite{maiti2008,usman2013}, which we attribute, as in KLT~\cite{kmcktl2011}, to the fact that the response of the dielectric matrix is not included, as well as its effect on the PNRs.

At low temperature, the hopping times $\tau (T;x) = \tau_0 e^{+ \Delta F_{loc}(T;x) / k_B T}$ are much larger than the period $t_e$: the block variables have not the time to follow the external field (no dielectric response). As T increases, $\tau (T;x)$ decreases, but the emerging macroscopic polarization exhibits a delay $\phi$ as long as $\tau (T;x) > t_e$. 
When $\tau (T;x) \sim t_e$, a {\it resonance} 
between the hopping times of the PNRs and the characteristic time of external solicitation yields the maximum of the response:
${\bf{P}}$ and the solicitation ${\bf{E_{ext}}}$ evolve in phase (the block variables have the time to follow the external field). Above $T_{max}$, the hopping times are so short that the block variables instantaneously, almost adiabatically, adapt their state to the external field: the dielectric response decreases owing to the thermal agitation, that tends to equalize the probabilities of the different minima.

$T_{max}$ therefore naturally depends on $\omega$, and the resonance condition $\tau (T_{max};x) \sim t_e$, i.e. $f = f_0 e^{- \Delta F_{loc}(T_{max};x) / k_B T_{max}}$ should provide the relation between $T_{max}$ and the external field frequency $f$. If there was a single relaxation time (one relaxor entity), with one single hopping barrier $U$ not dependent on $T$, as in KLT, this would simply provide $f = f_0 e^{- U / k_B T_{max}}$. Here, several relaxation times $\tau (T;x)$ coexist, associated to the existence of blocks with different chemical compositions $x$ and, moreover, interacting with each other (which changes the local barriers). The relation $f(T_{max})$ is thus complex, and mainly controlled by the form of $H_{loc}^{(N)}$. Having in mind that the phenomenological form chosen for $H_{loc}^{(N)}$ directly influences $f(T_{max})$, we can perform a fit of these data (i) on the Arrhenius law,  $ln f = ln f_0 - \frac{U}{k_B T_{max}}$, (ii) on the Vogel-F\"ulcher law, $ln f = ln f_0 - \frac{U}{k_B(T_{max}-T_f)}$, and (iii) on a relation of the form $ln f = ln f_0 - \alpha \frac{(T_{max}-T_1)^2}{T_{max}}$, reflecting the temperature evolution of the free energy barriers in our phenomenological model (Fig.~\ref{fig2}a).
We find the data compatible with (ii) and (iii), with $T_f$ = 127.6 K, $U$=0.019 eV and $f_0 \sim$ 1.0 $\times$ 10$^{10}$ Hz in case (ii), in reasonable agreement with experiments~\cite{dixit2006}, and $f_0$=2.5$\times$10$^{7}$ Hz, $\alpha$=1.89753 and $T_1$=167.3 K in case (iii).

Finally, we add that two kinds of relaxation are observed in BZT, the first one in the hyper frequency range, the second one in the radiofrequency range~\cite{petzelt2014,nuzhnyy2012}. While the classical MD simulations of Refs.~\cite{wang2011,wang2016} showed that the first relaxation is due to {\it single} Ti motions, our present method reveals that the second relaxation rather originates from {\it collective} motions of local modes inside PNRs, demonstrating the complementary of classical MD (that can mimic the faster individual hoppings, but not the slower collective dynamics in the whole PNRs) and our developed scheme (that can model such latter dynamics, as a result of our block compartments).

In summary, we have developed a methodology to simulate the radiofrequency dielectric response of relaxor materials. It is based on a RG transformation combined with KMC. 
We have shown that, in BZT, it allows modeling the collective motions of correlated local modes, and reproduces the radiofrequency dielectric relaxation for the first time ever with a potentially fully atomistic-based method. 
In particular, the local RFs related to the difference in polarizability between Ti and Zr ions are sufficient to explain radiofrequency relaxation, an important step toward a deep understanding of relaxors (non-local electric and elastic RFs are weak in BZT~\cite{bzt}, while they play an important role in heterovalent relaxors \cite{pmn} -- implying that they likely contribute to the radiofrequency response of PMN and similar systems). Our approach still needs to be tested on systems with larger correlation lengths (which will be the topic of a future study), for which theoretical difficulties have been pointed out in the past~\cite{pytte1981,imbrie1984}. As a matter of fact, in such systems, it may be more difficult to construct an approximate coarse-grained Hamiltonian as in Eq.~\ref{hn}. 
We hope that our approach will be applied and/or generalized to many complex materials and phenomena in the future.

\begin{acknowledgments}
L.B. acknowledges ONR Grant N00014-12-1-1034.
\end{acknowledgments}

\newpage

\begin{equation}
\nonumber
Supplemental~~information
\end{equation}

\section{Introduction}

Theoretical and technical details are provided hereafter. 

We first recall the main ideas of our methodology to study relaxors. 
Our approach starts from the search for some local description of the free energy landscape in a relaxor compound, as a function of a set of local order parameters, defined at the scale of the polar nano-regions (PNRs). From a more fundamental viewpoint, it is inspired from Real-Space Renormalization Group (RG) theories~\cite{introcritical}. Basically, these methods consist in reducing the number of degrees of freedom of a system by gathering them in blocks ($N \times N \times N$), retaining only one variable per block ("block variable") and defining a new hamiltonian that preserves the physical properties, i.e. the partition function. 
For that, the partition function is incompletely calculated by integrating over all the degrees of freedom except the block variables, which are maintained fixed, providing the possibility of defining (up to an additive constant), this new Hamiltonian $H^{(N)}$ ("coarse-grained Hamiltonian"). $H^{(N)}$ is a function of these block variables, and has the same partition function (at least up to a multiplicative constant) as the initial one. It depends on the temperature and is an {\it incomplete free energy}~\cite{geneste2011,geneste2009,geneste2010}.

Such transformation increases the length scale of the system from $a_0$ to $N a_0$ ($a_0$ being the lattice constant of the initial lattice). The new system, having the same partition function as the starting one, has exactly the same static observables: it is {\it macroscopically} indistinguishable, and corresponds to a statistical description of the system at a new length scale, namely $N a_0$. The coarse-grained hamiltonian (coarse-grained free energy), its precise definition, construction and approximation, are detailed hereafter.

Renormalization group transformations have been developed to describe the behavior of a system in the vicinity of second-order phase transitions, the so-called "critical phenomena". In such systems (e.g., magnetic materials), the microscopic degrees of freedom correlate on larger and larger distances as the phase transition is approached. This is associated with a divergence of the correlation length at the transition point and a slowing down of the dynamics ("critical slowing down"). The formation of large structures with slow dynamics is a common point that such systems share with relaxors. It is thus quite natural to apply some concepts of the renormalization group to relaxors. 
In real-space RG transformations, the new lattice is eventually contracted by a factor $N$ to map onto the initial one, allowing to iterate the transformation~\cite{introcritical}. In our case, however, the transformation will not be iterated, because typical length scales do not diverge in a relaxor. We will only use the RG transformation to reach the length scale of these structures, and will keep the size of the blocks lower than the correlation length, in order to make possible some approximate analytical calculation of $H^{(N)}$.

We first detail the construction of the coarse-grained Hamiltonian, and the way it is given an approximate form.
Then we describe how each of its parts can be computed, and explain how to make the link between the thermodynamic aspects and the kinetic ones. Finally, technical details are given about the Kinetic Monte Carlo simulations.

Our study is performed in the framework of {\it Classical Statistical Physics}.
We will use the following notations:

\begin{itemize}

\item $\{ {\bf u_i^{(1)}}  \}$ = $\{  {\bf u_1^{(1)}}, {\bf u_2^{(1)}}, {\bf u_3^{(1)}} ...  \}$ is the set of {\it all} the microscopic local modes.

\item $\{ {\bf u_k^{(N)}}  \}$ = $\{  {\bf u_1^{(N)}}, {\bf u_2^{(N)}}, {\bf u_3^{(N)}} ...  \}$ is the set of {\it all} the block variables.

\item   ${\bf v }. \bar{\bar{C}} . {\bf u } = \sum_{\alpha,\beta} \bar{\bar{C}}_{\alpha,\beta} v_{\alpha} u_{\beta}$, in which $\bar{\bar{C}}_{\alpha,\beta}$ is a symmetric 3 $\times$ 3 matrix. $\alpha$ and $\beta$ run over the cartesian directions ($X$, $Y$, $Z$).

\item $< ... >$  =  thermal average, i.e. in the canonical ensemble:

\begin{eqnarray}
\nonumber
<A> = C \int ... \int  A(\{ {\bf u_i^{(1)} } \})
\frac{e^{- \beta H^{(1)}(\{ {\bf u_i^{(1)} } \})}}{Z} \{ \prod_{i} d {\bf u_i^{(1)} } \} ,
\end{eqnarray}

 with 

\begin{eqnarray}
\nonumber
Z = C \int ... \int e^{- \beta H^{(1)}(\{ {\bf u_i^{(1)}} \})} \{ \prod_{i} d {\bf u_i^{(1)}} \} ,
\end{eqnarray}

the canonical partition function, and $\beta = \frac{1}{k_B T}$.
We have assumed that $A$ is an observable depending only on the configuration variables ${\bf u_i^{(1)}}$, not on their conjugate momenta ${\bf p_i^{(1)}}$ (which have been integrated out in the constant $C$ in the expression of Z and $<A>$). 
$H^{(1)}$ is, by convention, the potential energy of the microscopic hamiltonian, i.e. it depends only on the configuration variables:
$H^{(1)}(\{  {\bf u_i^{(1)} } \})$.

\item  $< ... >({\bf u})$  =  thermal average under fixed ${\bf u} = \frac{1}{P} \sum_{i=1}^P {\bf u_i^{(1)}}$

\begin{eqnarray}
\nonumber
<A>({\bf u}) = C \int ... \int  A(\{ {\bf u_i^{(1)} } \}) \delta(P {\bf u } - \sum_{i=1}^P {\bf u_i^{(1)} })
\frac{e^{- \beta H^{(1)}(\{ {\bf u_i^{(1)} } \})}}{\tilde{Z}({\bf u})} \{ \prod_{i} d {\bf u_i^{(1)} }\} ,
\end{eqnarray}

 with 

\begin{eqnarray}
\nonumber
\tilde{Z}({\bf u }) = C \int ... \int \delta(P {\bf u } - \sum_{i=1}^P {\bf u_i^{(1)} })
e^{- \beta H^{(1)}(\{ {\bf u_i^{(1)} } \})} \{ \prod_{i} d {\bf u_i^{(1)} } \} 
\end{eqnarray}

$< A >({\bf u})$ is a conditional average~\cite{sprik1998} evaluated for $\frac{1}{P} \sum_{i=1}^P {\bf u_i^{(1)}} = {\bf u}$.

\end{itemize}

\section{The Microscopic Hamiltonian}

Our microscopic description of relaxors is based on the concept of effective Hamiltonian~\cite{zhong1995,zhong1994}.

\subsection{Effective Hamiltonian without chemical disorder}

The degrees of freedom of the perovskite lattice are reduced to the local modes $\{ {\bf u_i^{(1)} } \}$, related to the electric dipoles that locally exist in each cell $i$, the homogeneous strain tensor $\{  \eta_l^H  \}$, the local mechanical displacement modes $\{ {\bf v_i^{(1)} } \}$, related to a possible local inhomogeneous strain at cell $i$, $\{  \eta_l^I(i)  \}$, and possibly other degrees of freedom when they appear as relevant, such as the antiferrodistortive modes ${\bf \omega_i^{(1)} }$ (for perovskite lattices with antiferrodistortions~\cite{afd}) or the local magnetic moments $\{ {\bf m_i }  \}$ (for multiferroic systems~\cite{bfo}). The total strain at cell $i$ is $\eta_l(i) = \eta_l^H + \eta_l^I(i)$. Standard ferroelectric systems such as BaTiO$_3$ (BTO) are well described by accounting for the $\{ {\bf u_i^{(1)} } \}$ and the $\{  \eta_l(i)  \}$. 

In the present work, we only account for the polar degrees of freedom, namely the local modes $\{ {\bf u_i^{(1)} }\}$, and consider a microscopic Hamiltonian $H^{(1)}$ having the following simplified form

\begin{eqnarray}
\nonumber
\label{heff}
H^{(1)}(\{ {\bf u_i^{(1)} } \})
=\sum_{i} H_{loc}^{(1)}({\bf u_i^{(1)} }) 
+ \frac{1}{2}  \sum_{i,j \atop (i \ne j)} {\bf u_i^{(1)} }. \bar{\bar{C}}_{SR}^{(1)}(i,j). {\bf u_j^{(1)} } 
+ \frac{1}{2}  \sum_{i,j \atop (i \ne j)} {\bf u_i^{(1)} }. \bar{\bar{C}}_{LR}^{(1)}(i,j). {\bf u_j^{(1)} },
\end{eqnarray}

in which ${\bar{\bar{C}}}_{LR,\alpha \beta}^{(1)}(i,j)$ and ${\bar{\bar{C}}}_{SR,\alpha \beta}^{(1)}(i,j)$ are the matrix elements of the long-range (LR) dipole-dipole and Short-Range (SR) interactions between cell $i$ and cell $j$.

The local part has the form 
\begin{eqnarray}
\nonumber
H_{loc}^{(1)}({\bf u }) = H_{loc}^{(1)}({\bf 0}) + \kappa_2^{(1)}  (u_X^2 + u_Y^2 + u_Z^2) +
\alpha^{(1)} || {\bf u} ||^4 + 
\gamma^{(1)} (u_X^2 u_Y^2 + u_X^2 u_Z^2 + u_Y^2 u_Z^2)
\end{eqnarray}

(In $H^{(1)}$, $H_{loc}^{(1)}({\bf 0})$=0, but we write it for consistency with the coarse-grained hamiltonian $H^{(N)}$).

\subsection{Effective Hamiltonian with chemical disorder, random fields}

In the case of relaxors, that exhibit chemical disorder on one of their sublattice, we associate to each unit cell $i$ a real number $x_i^{(1)}$, that characterizes the local composition of the cell.

The dependence on $x_i^{(1)}$ of any term in the hamiltonian characterizes the presence of "random fields". However, we distinguish {\it local random fields} from {\it non-local random fields}, whether the dependence on $x_i^{(1)}$ appears in the local term or in the interaction terms. 
The presence of at least one kind of random field, either local or non-local, is of course necessary to have a relaxor behavior associated to the presence of polar nanoregions (otherwise, the hamiltonian is simply that of a ferroelectric system).

In the case of Ba(Zr$_{0.5}$Ti$_{0.5}$)O$_3$ (BZT), we choose $x_i^{(1)}$=0 (resp. 1) if the chemical species is Zr (resp. Ti). The $\{ x_i^{(1)} \}$ are parameters (not variables) of the effective Hamiltonian. We use the following form

\begin{eqnarray}
\label{heff2}
\nonumber
H^{(1)}(\{ {\bf u_i^{(1)} } \};\{x_i^{(1)} \})=
\underbrace{\sum_{i} H_{loc}^{(1)}({\bf u_i^{(1)} } ; x_i^{(1)})}_{local~random~fields}
+ \frac{1}{2} \sum_{i,j \atop (i \ne j)} {\bf u_i^{(1)} }. \bar{\bar{C}}_{SR}^{(1)}(i,j). {\bf u_j^{(1)} }
+ \frac{1}{2} \sum_{i,j \atop (i \ne j)} {\bf u_i^{(1)} }. \bar{\bar{C}}_{LR}^{(1)}(i,j). {\bf u_j^{(1)} },
\end{eqnarray}

in which the chemical disorder only contributes to the local part:

\begin{eqnarray}
\nonumber
H_{loc}^{(1)}({\bf u } ; x) = H_{loc}^{(1)}({\bf 0}) + \kappa_2^{(1)}(x)  (u_X^2 + u_Y^2 + u_Z^2) +
\alpha^{(1)}(x) || {\bf u} ||^4 + 
\gamma^{(1)}(x) (u_X^2 u_Y^2 + u_X^2 u_Z^2 + u_Y^2 u_Z^2)
\end{eqnarray}

This Hamiltonian is an approximation of the one for BZT described in Ref.~\onlinecite{bzt}, and contains thus, according to the previous definition, only {\it local random fields}.
By contrast, "non-local random fields" would be characterized by terms of the kind 
$\sum_{i,j \atop (i \ne j)} {\bf u_i^{(1)} }. \bar{\bar{C}}_{LR}^{(1)}(i,j)[x_i^{(1)},x_j^{(1)}]. {\bf u_j^{(1)} }$.
Such terms are present in the hamiltonian of heterovalent relaxors such as PbMg$_{1/3}$Nb$_{2/3}$O$_3$ (PMN), because the two cations do not bear the same charge.

\section{The coarse-grained Hamiltonian}

\subsection{The coarse-graining approach}

\subsubsection{Definition of the coarse-grained Hamiltonian}
The unit cells are gathered by blocks of size $N \times N \times N$ lattice constants. A block variable is defined in each block $k$, as
\begin{equation}
{\bf u_k^{(N)} } = \frac{1}{N^3} \sum_{i \in k} {\bf u_i^{(1)} }
\end{equation}

In the following, we use the term "block variable" or "local order parameter" to denote the $\{ {\bf u_k^{(N)} }  \}$. To make the formulas more readable, we will use $i$ as the index running over the elementary unit cells, and $k$ as the index running over the blocks.

In these three first subsections, we ignore chemical disorder.
The set of local order parameters $\{ {\bf u_k^{(N)} }  \}$ is used to define an incomplete partition function~\cite{binder1987,geneste2011}, 

\begin{eqnarray}
\nonumber
\tilde{Z}^{(N)}(\{ {\bf u_k^{(N)} } \})  = 
C \int ... \int \{ \prod_{blocks~k} \delta(\sum_{i \in k} {\bf u_i^{(1)} }-N^3 {\bf u_k^{(N)} }) \}
e^{- \beta H^{(1)}(\{ {\bf u_i^{(1)} } \})} \{ \prod_{i} d {\bf u_i^{(1)} } \} ,
\end{eqnarray}

by summing over the microscopic states $\{ {\bf u_i^{(1)} } \}$ such that $\forall~k$, $\sum_{i \in k} {\bf u_i^{(1)} } = N^3 {\bf u_k^{(N)} }$ ($\delta$ is the Dirac function), i.e. the summation is restricted to the microscopic states having as block variables the set of values $\{ {\bf u_k^{(N)} } \}$.

The $\{ {\bf u_k^{(N)} } \}$ are thus variables of this incomplete partition function. 
The corresponding coarse-grained Hamiltonian is defined, up to an additive constant, as
\begin{equation}
\label{cgdef}
H^{(N)}(\{ {\bf u_k^{(N)} }  \}) = - k_B T ln  \tilde{Z}^{(N)}(\{ {\bf u_k^{(N)} } \})
\end{equation}

A few remarks concerning $H^{(N)}$:

\begin{itemize}

\item the notion of coarse-grained Hamiltonian is a basic concept of Real-Space Renormalization Group transformations~\cite{binder1987,introcritical};

\item the coarse-grained Hamiltonian is an {\bf incomplete free energy}, because it is built from a summation of terms proportional to the canonical probability over a subset of phase space~\cite{free-en1,free-en2};

\item by contrast with $H^{(1)}$, the coarse-grained Hamiltonian $H^{(N)}$ depends on the temperature, by construction, owing to the renormalization process~\cite{introcritical} previously described;

\item $H^{(N)}$ is expected to be an extensive quantity; this is reflected by the formula providing its derivatives (see next subsection, Eq.~\ref{meanforce}).

\item by construction, $H^{(N)}$ and $H^{(1)}$ have obviously the same partition function (at least up to a multiplicative constant). The system of the $\{  {\bf u_i^{(1)} }  \}$ interacting through $H^{(1)}$, and that of the $\{  {\bf u_k^{(N)} } \}$ interacting through $H^{(N)}$ have therefore the same static observables, and are macroscopically identical.

\item the density of probability of a set of values for the block variables is $P(\{ {\bf u_k^{(N)} }  \}) = \frac{\tilde{Z}^{(N)} (\{ {\bf u_k^{(N)} }  \})}{Z}$. 

\item $H^{(N)}(\{ {\bf u_k^{(N)} }  \})$ is defined from Eq.~\ref{cgdef} {\it up to an additive constant} (this is reflected in the fact that $ \tilde{Z}^{(N)}$ in the logarithm has a physical dimension). Alternatively, $H^{(N)}$ can be defined as

\begin{equation}
\nonumber
H^{(N)}(\{ {\bf u_k^{(N)} }  \}) - H^{(N)}(\{ {\bf u_k^{(N)} } = {\bf 0}  \})
= - k_B T ln \frac {  \tilde{Z}^{(N)}(\{ {\bf u_k^{(N)} } \}) }  { \tilde{Z}^{(N)}(\{ {\bf u_k^{(N)} }  = {\bf 0}\}) }   ,
\end{equation}

The additive constant has no influence on the following results, because in the Kinetic Monte Carlo, only free energy barriers are considered, i.e. differences of $H^{(N)}$.

\end{itemize}

\subsubsection{Derivative of the coarse-grained Hamiltonian}
\label{mean}

To calculate the coarse-grained free energy $H^{(N)}$, we use the following formula~\cite{geneste2009,geneste2010,geneste2011}:

\begin{equation}
\label{meanforce}
\forall n,~~
\frac{\partial H^{(N)}}{\partial {\bf u_n^{(N)} } } (\{ {\bf u_k^{(N)} } \}) = - < \sum_{i \in n} {\bf f_i } > (\{ {\bf u_k^{(N)} } \}),
\end{equation}

in which ${\bf f_i }$ is the microscopic force acting on local mode $i$, i.e. ${\bf f_i } = - \frac{\partial H^{(1)}}{\partial {\bf u_i^{(1)} }}$. It states that the partial derivative of $H^{(N)}$ with respect to ${\bf  u_n^{(N)} }$ is (minus) the mean total force in block $n$, {\it under the constraint of fixed block variables} $\{ {\bf u_k^{(N)} } \}$.
Similar formulas have already been used in Refs.~\onlinecite{geneste2009,geneste2010,geneste2011} to compute free energies of ferroelectric systems as a function of polarization. $H^{(N)}$ can be viewed as a potential of mean force, as currently used in chemical physics to study chemical processes along a reaction coordinate~\cite{sprik1998,free-en1,free-en2}. In the present case, the "reaction coordinate" is multidimensional: it is the set of all the block variables. 

We now provide a proof for Eq.~\ref{meanforce}:
$\frac{\partial H^{(N)}}{\partial {\bf  u_n^{(N)} }} (\{ {\bf u_k^{(N)} } \}) = - < \sum_{i \in n} {\bf f_i } > (\{ {\bf u_k^{(N)} } \})$

We start from the incomplete partition function, subject to a given set of values for the block variables:

\begin{eqnarray}
\nonumber
\tilde{Z}^{(N)}(\{ {\bf u_k^{(N)} } \}) = C \int ... \int \{ \prod_{blocks~k} \delta(\sum_{i \in k} {\bf u_i^{(1)} }-N^3 {\bf u_k^{(N)} }) \}
e^{- \beta H^{(1)}(\{ {\bf u_i^{(1)} } \})} \{ \prod_{i} d {\bf u_i^{(1)} } \} 
\end{eqnarray}


We select one block $n$, and calculate the derivative $\frac{\partial H^{(N)}}{\partial {\bf u_n^{(N)} }}$:

\begin{equation}
\frac{\partial H^{(N)}}{\partial {\bf  u_n^{(N)} }} = - \frac{ k_B T }{\tilde{Z}^{(N)}}   
\frac{\partial \tilde{Z}^{(N)}}{\partial {\bf u_n^{(N)} }}
\end{equation}

We have thus to derive $\tilde{Z}^{(N)}$ with respect to ${\bf u_n^{(N)} }$. For that, we modify the expression of $\tilde{Z}^{(N)}$ by replacing in the integral
one of the ${\bf u_i^{(1)} }$ of block $n$, let us call it ${\bf u_I^{(1)} }$, by $N^3 {\bf u_n^{(N)} } - \sum_{i \ne I \atop i \in n} {\bf u_i^{(1)} }$ and eliminating the corresponding $\delta$ function from the product $\Pi_k$, and $d {\bf u_I^{(1)} }$ from the integral:

\begin{eqnarray}
\nonumber
\tilde{Z}^{(N)}(\{ {\bf u_k^{(N)} } \}) = C \int ... \int \{ \prod_{blocks~k \ne n} \delta(\sum_{i \in k} {\bf u_i^{(1)} }-N^3 {\bf u_k^{(N)} }) \}
 e^{- \beta H^{(1)}(\{ {\bf u_i^{(1)} }, i \ne I \}, {\bf u_I^{(1)} }= N^3 {\bf u_n^{(N)} } - \sum_{i \ne I \atop i \in n} {\bf u_i^{(1)} })} 
\{ \prod_{i \ne I} d {\bf u_i^{(1)} } \}
\end{eqnarray}

In this expression, the block variable ${\bf u_n^{(N)} }$ appears now only in the Boltzmann term. We can thus easily derive:

\begin{eqnarray}
\nonumber
\frac{\partial \tilde{Z}^{(N)}}{\partial {\bf u_n^{(N)} }} (\{ {\bf u_k^{(N)} } \}) =
C \int ... \int    \{ 
\prod_{blocks~k \ne n} \delta(\sum_{i \in k} {\bf u_i^{(1)} }-N^3 {\bf u_k^{(N)} }) \}  \\
\times  \{  - \beta N^3 \frac{\partial H^{(1)}}{\partial {\bf u_I^{(1)} } }  \} 
 e^{- \beta H^{(1)}(\{ {\bf u_i^{(1)} }, i \ne I \}, {\bf u_I^{(1)} }= N^3 {\bf u_n^{(N)} } - \sum_{i \ne I \atop i \in n} {\bf u_i^{(1)} })} 
\{ \prod_{i \ne I} d {\bf u_i^{(1)} } \}  
\end{eqnarray}

Dividing by ${\tilde{Z}^{(N)}}(\{ {\bf u_k^{(N)} } \})$, multiplying by $- k_B T$ and restauring the integration over ${\bf u_I^{(1)} }$, we have

\begin{eqnarray}
\nonumber
\frac{\partial H^{(N)}}{\partial {\bf  u_n^{(N)} }} =
C \int ... \int    
[  N^3 \frac{\partial H^{(1)}}{\partial {\bf u_I^{(1)} } }  ]
\{ 
\prod_{blocks~k} \delta(\sum_{i \in k} {\bf u_i^{(1)} }-N^3 {\bf u_k^{(N)} }) \}  
\frac{
 e^{- \beta H^{(1)}(\{ {\bf u_i^{(1)} } \}  )} } 
{{\tilde{Z}^{(N)} (\{ {\bf u_k^{(N)} } \})}} 
\{ \prod_{i} d {\bf u_i^{(1)} } \} 
\end{eqnarray}

and the previous expression appears as the thermal average under the constraint of fixed $\{ {\bf u_k^{(N)} } \}$ of the force on the local mode $I$

\begin{equation}
\frac{\partial H^{(N)}}{\partial {\bf  u_n^{(N)} }}(\{ {\bf u_k^{(N)} } \}) = - N^3 < {\bf f_I } > (\{ {\bf u_k^{(N)} } \})
\end{equation}

Since this expression is valid for all $I$ belonging to block $n$, we have

\begin{equation}
\frac{\partial H^{(N)}}{\partial {\bf u_n^{(N)} }}(\{ {\bf u_k^{(N)} } \}) = -  < \sum_{i \in n}  {\bf f_i } > (\{ {\bf u_k^{(N)} }\})
\end{equation}

\subsubsection{Approximate form for the coarse-grained Hamiltonian}
\label{approx}
We now explain how this coarse-grained Hamiltonian can be given an approximate functional form identical to that of the microscopic Hamiltonian $H^{(1)}$ (following the spirit of Real-Space Renormalization Group techniques~\cite{introcritical}), i.e.

\begin{eqnarray}
\label{hn}
\nonumber
H^{(N)}  (\{ {\bf u_k^{(N)} } \}) \approx
\sum_{n} H_{loc}^{(N)} ( {\bf u_n^{(N)} } ) + 
\frac{1}{2} \sum_{n,n' \atop (n \ne n')} 
{\bf u_n^{(N)} }. \bar{\bar{C}}_{SR}^{(N)} (n,n'). {\bf u_{n'}^{(N)} } +
\frac{1}{2} \sum_{n,n' \atop (n \ne n')} 
{\bf u_n^{(N)} }. \bar{\bar{C}}_{LR}^{(N)} (n,n'). {\bf u_{n'}^{(N)} },
\end{eqnarray}

and under which assumptions this is possible.

To obtain such form, we start from Eq.~\ref{meanforce}: the force  ${\bf f_i }$ on the local mode $i$ writes

\begin{eqnarray}
{\bf f_i } = - {\frac{\partial H^{(1)}}{\partial {\bf u_i^{(1)} } }} =
- \frac{\partial H_{loc}^{(1)}}{\partial {\bf u}} ({\bf u_i^{(1)} }) 
- \sum_{j, j \ne i} \bar{\bar{C}}^{(1)}(i,j) . {\bf u_j^{(1)} },
\end{eqnarray}

with $\bar{\bar{C}}^{(1)}(i,j) = \bar{\bar{C}}^{(1)}_{SR}(i,j) + \bar{\bar{C}}^{(1)}_{LR}(i,j)$.

Summing over $i \in n$, we have

\begin{eqnarray}
\sum_{i \in n} {\bf f_i } = 
- \sum_{i \in n} \frac{\partial H_{loc}^{(1)}}{\partial {\bf u }} ({\bf u_i^{(1)} })
- \sum_{i \in n, j \atop i \ne j} \bar{\bar{C}}^{(1)}(i,j) . {\bf u_j^{(1)} } 
\end{eqnarray}

We now make the thermal average under fixed $(\{ {\bf u_k^{(N)} } \})$

\begin{eqnarray}
\nonumber
< \sum_{i \in n} {\bf f_i } > (\{ {\bf u_k^{(N)} } \}) = 
- \sum_{i \in n} <\frac{\partial H_{loc}^{(1)}}{\partial {\bf u }} ({\bf u_i^{(1)} })> (\{ {\bf u_k^{(N)} } \}) 
- \sum_{i \in n, j \atop i \ne j} \bar{\bar{C}}^{(1)}(i,j) . <{\bf u_j^{(1)} }> (\{ {\bf u_k^{(N)} } \}) ,
\end{eqnarray}

and separate $\sum_j$ in contributions coming from the different blocks, $n$ and $n' \ne n$:

\begin{equation}
\sum_j = \sum_{j \in n} + \sum_{j \notin n} = \sum_{j \in n} + \sum_{n' \atop (n'\ne n)} \sum_{j \in n'}
\end{equation}

Finally, the derivative of $H^{(N)}$ writes

\begin{eqnarray}
\nonumber
\frac{ \partial H^{(N)} }{\partial {\bf u_n^{(N)} }}  (\{ {\bf u_k^{(N)} } \}) =
- < \sum_{i \in n} {\bf f_i } > (\{ {\bf u_k^{(N)} } \}) =  \\
\nonumber
\sum_{i \in n} 
\{
<\frac{\partial H_{loc}^{(1)}}{\partial {\bf u }} ({\bf u_i^{(1)} })> (\{ {\bf u_k^{(N)} } \}) 
+ \sum_{j \in n \atop j \ne i} \bar{\bar{C}}^{(1)}(i,j).<{\bf u_j^{(1)} }> (\{ {\bf u_k^{(N)} } \})
\}  \\
\nonumber
 + \sum_{n' \atop (n'\ne n)} \{
\sum_{i \in n \atop j \in n'} \bar{\bar{C}}_{SR}^{(1)}(i,j) . <{\bf u_j^{(1)} }> (\{ {\bf u_k^{(N)} } \})
\}  \\
\label{derivative1}
 + \sum_{n' \atop (n'\ne n)} \{
\sum_{i \in n \atop j \in n'} \bar{\bar{C}}_{LR}^{(1)}(i,j) . <{\bf u_j^{(1)} }> (\{ {\bf u_k^{(N)} } \})
\}  
\end{eqnarray}

Within this last expression, the derivative of $H^{(N)}$ with respect to ${\bf u_n^{(N)} }$ appears as consisting of three parts: (i) a local contribution from block $n$, (ii) a sum of contributions from all the other block $n'$ involving the short-range interaction, and (iii) a sum of contributions from all the other block $n'$ involving the long-range interaction.

Up to this stage, our derivation is exact. 
Of course, it would be convenient to have an analytical expression for $H^{(N)}$, but this is not possible in general.
To go further, we have to make some approximations, based on two assumptions:

\begin{enumerate}

\item Local Homogeneity: we make the approximation that the average value of the local mode ${\bf u_i^{(1)} }$ in cell $i$ 
under the constraint of fixed block variables $\{ {\bf u_k^{(N)} } \}$,
is equal to the mean value of the block to which it belongs.
$\forall i \in n, < {\bf u_i^{(1)} } >(\{ {\bf u_k^{(N)} } \}) = {\bf u_n^{(N)} }$.

\item We make the supplementary approximation that the average under fixed $\{{\bf u_k^{(N)} } \}$ 
of any function in the $u_{i,\alpha}^{(1)}$, of the form  $< {u_{i,\alpha}^{(1)}}^2 u_{i,\beta}^{(1)} > (\{ {\bf u_k^{(N)} } \})$, etc ... 
is a function of the local block variable to which $i$ belongs only (and not of that of the neighboring blocks):

$\forall i \in n,
< {u_{i,\alpha}^{(1)}}^2 u_{i,\beta}^{(1)} > (\{ {\bf u_k^{(N)} } \}) = 
< {u_{i,\alpha}^{(1)}}^2 u_{i,\beta}^{(1)} > ( {\bf u_n^{(N)} } )$

It also depends naturally on the temperature.
These functions appear in the mean force when deriving the local part of $H^{(1)}$.

\end{enumerate}

These approximations imply blocks of size lower than the correlation length ($N a_0 \leq \xi$)~\cite{geneste2010,troster2005} (otherwise, a block can demix into domains, see Ref.~\onlinecite{troster2005}). 
Note that assumption (1) need not be true for all values of the block variables: it is sufficient that it stands for ${\bf u_n^{(N)} }$ around the local minima and along the transition paths between local minima. 
However, if it is reasonable to assume (1) as true around local minima of $H^{(N)}$, its validity at the saddle point on the transition path between two minima (thus, typically for ${\bf u_n^{(N)} }$ along $<$110$>$ directions) is not that obvious: we have to consider the reversal of the block variables as slow enough so that the states under the constraint of fixed ${\bf u_n^{(N)} }$ are ergodic along the transition path.

Under these two approximations, we have

\begin{eqnarray}
\nonumber
\frac{ \partial H^{(N)} }{\partial {\bf u_n^{(N)} }}  (\{ {\bf u_k^{(N)} } \}) \approx 
\sum_{i \in n} 
\{
<\frac{\partial H_{loc}^{(1)}}{\partial {\bf u }} ({\bf u_i^{(1)} })> ( {\bf u_n^{(N)} } ) +
 \sum_{j \in n \atop j \ne i} \bar{\bar{C}}^{(1)}(i,j). {\bf u_n^{(N)} }
\}  +  
\sum_{n' \atop (n'\ne n)} \{
\sum_{i \in n \atop j \in n'} \bar{\bar{C}}^{(1)}(i,j)  \} . {\bf u_{n'}^{(N)} } 
\end{eqnarray}

Integration of $\frac{ \partial H^{(N)} }{\partial {\bf u_n^{(N)} }} $ provides the following form

\begin{eqnarray}
\label{hnbis}
H^{(N)}  (\{ {\bf u_k^{(N)} } \}) \approx
\sum_{n} H_{loc}^{(N)} ({\bf u_n^{(N)} }) +  
\frac{1}{2} \sum_{n,n' \atop (n' \ne n)} 
{\bf  u_n^{(N)} }. \bar{\bar{C}}^{(N)} (n,n'). {\bf u_{n'}^{(N)} }
\end{eqnarray}

with
\begin{equation}
\bar{\bar{C}}^{(N)} (n,n')=\bar{\bar{C}}_{SR}^{(N)} (n,n')+\bar{\bar{C}}_{LR}^{(N)} (n,n')
\end{equation}

\begin{equation}
\bar{\bar{C}}_{SR}^{(N)} (n,n')=
\sum_{i \in n} \sum_{j \in n'} \bar{\bar{C}}_{SR}^{(1)} (i,j)
\end{equation}

\begin{equation}
\bar{\bar{C}}_{LR}^{(N)} (n,n')=
\sum_{i \in n} \sum_{j \in n'} \bar{\bar{C}}_{LR}^{(1)} (i,j)
\end{equation}

and

\begin{eqnarray}
\label{integration-thermo}
\frac{\partial H_{loc}^{(N)}}{\partial {\bf u }} ({\bf u_n^{(N)} }) =
\sum_{i \in n} 
\{
<\frac{\partial H_{loc}^{(1)}}{\partial {\bf u }} ({\bf u_i^{(1)} })> ( {\bf u_n^{(N)} } ) + 
 \sum_{j \in n \atop j \ne i} [\bar{\bar{C}}_{SR}^{(1)}(i,j)+\bar{\bar{C}}_{LR}^{(1)}(i,j)]. {\bf u_n^{(N)} }
\} 
\end{eqnarray}

In Eq.~\ref{hnbis}, $H^{(N)}$ has exactly the same functional form as $H^{(1)}$. One can push one step further the similarity between $H^{(1)}$ and $H^{(N)}$, since in the domain of validity of our approximations, $H_{loc}^{(N)}$ is likely to have a smooth and continuous evolution, and can be expanded in power series of ${\bf u}$. If this Landau expansion of the local free energy is truncated to fourth-order, as in the Landau theory of second-order phase transitions, even $H_{loc}^{(N)}$ can be enforced in the same functional form as $H_{loc}^{(1)}$:

\begin{eqnarray}
\nonumber
H_{loc}^{(N)}({\bf u }) = H_{loc}^{(N)}({\bf 0}) + \kappa_2^{(N)} (T) (u_X^2 + u_Y^2 + u_Z^2) +  
\alpha^{(N)}(T) || {\bf u} ||^4 +  
\gamma^{(N)}(T) (u_X^2 u_Y^2 + u_X^2 u_Z^2 + u_Y^2 u_Z^2)
\end{eqnarray}

By the way, it is indeed commonly believed that coarse-grained Hamiltonians have the Landau form around the critical temperature for systems undergoing a second-order phase transition~\cite{binder1987,geneste2010}. However, 
the renormalized local free energy $H_{loc}^{(N)}$ could be ideally obtained by thermodynamic integration methods inside each block, following the spirit of Refs.~\onlinecite{geneste2009,geneste2010,geneste2011,kumar2010}. 
The renormalized SR and LR inter-block matrices write

\begin{equation}
\bar{\bar{C}}_{SR}^{(N)} (n,n')=
\sum_{i \in n} \sum_{j \in n'} \bar{\bar{C}}_{SR}^{(1)} (i,j)
\end{equation}

and

\begin{equation}
\bar{\bar{C}}_{LR}^{(N)} (n,n')=
\sum_{i \in n} \sum_{j \in n'} \bar{\bar{C}}_{LR}^{(1)} (i,j)
\end{equation}

The assumption that the local modes are sufficiently correlated over the blocks is absolutely fundamental. If this condition is not fulfilled, the renormalized form written above cannot be justified.

\subsubsection{Extension to the case of homovalent relaxors}
\label{relaxors}

We now apply the previous formalism to a certain family of relaxors. Relaxors with perovskite structure are characterized by a chemical disorder on one sublattice, involving at least two chemical species. This disorder can be homovalent, for instance in BZT, or heterovalent, for instance in PMN. In homovalent systems, the two chemical species involved in the disorder have the same formal charge (Ti$^{4+}$ and Zr$^{4+}$ in the case of BZT). This allows the composition to deviate locally from its nominal value without creating space-charge regions. According to Ref.~\onlinecite{bzt}, the formation of polar nano-regions in BZT is  related to such local deviation, since PNRs seem to be regions in which the Ti concentration is somewhat higher than in the rest of the matrix~\cite{bzt}. We limit the present study to this class of systems.

We consider a chemical disorder associated with two chemical species A and B on one sublattice of the perovskite network. Let us denote by $\delta$ the nominal composition of the system in terms of the B species, namely $A_{1 - \delta} B_{\delta}$, 0 $\leq \delta \leq$ 1. Among the two species A and B, one of them drives the polar instability, we will assume this is B (B=Ti, A=Zr in the case of BZT). We associate to each unit cell $i$ a real number $x_i^{(1)}$ that characterizes its local composition. At the scale of one single cell, $x_i^{(1)}$=0 (if the chemical species is A) or 1 (if the chemical species is B). This set of real numbers $\{ x_i^{(1)} \}$ plays the role of a set of parameters (not variables) of the effective Hamiltonian (in any simulation, they are fixed once for all at the beginning and do not change).
The probability to have $A$ (resp. $B$) in a given unit cell is $1 - \delta$ (resp. $\delta$). It is possible to associate a random variable $X_i^{(1)}$ to each unit cell $i$. This random variable characterizes the chemical disorder in the materials. 
The set of real numbers $\{ x_i^{(1)} \}$ is an occurrence of the set of random variables $\{ X_i^{(1)} \}$.

As explained above, we use the following form of the effective Hamiltonian, in which the chemical disorder contributes only to the local part:

\begin{eqnarray}
\label{heff2}
\nonumber
H^{(1)}(\{ {\bf u_i^{(1)} }\};\{x_i^{(1)} \})=
\sum_{i} H_{loc}^{(1)}({\bf u_i^{(1)} };x_i^{(1)}) 
+ \frac{1}{2} \sum_{i \ne j} {\bf u_i^{(1)} }. \bar{\bar{C}}^{(1)}(i,j). {\bf u_j^{(1)} } 
\end{eqnarray}

The presence of $x_i^{(1)}$ in the local part of $H^{(1)}$ only, not in the interaction terms, corresponds to the absence of non-local random field in the microscopic description of the system.

As in the previous section, we now gather the unit cells by blocks of size $N \times N \times N$. The set of these blocks forms a partition of the whole system. In block $k$, the local order parameter (block variable) is still defined as

\begin{equation}
\nonumber
{\bf u_k^{(N)} } = \frac{1}{N^3} \sum_{i \in k} {\bf u_i^{(1)} }
\end{equation}

This block variable ${\bf u_k^{(N)} }$ is proportional to the mean electric dipole in block $k$ because in the homovalent relaxor we consider, the effective charge is the same for Ti and for Zr.

We introduce another number $x_k^{(N)}$ that characterizes the local chemical disorder in block $k$:

\begin{equation}
\nonumber
x_k^{(N)} = \frac{1}{N^3} \sum_{i \in k} x_i^{(1)}
\end{equation}

$\{ x_k^{(N)} \}$ is a set of real numbers which takes fractional values in the interval [0;1]. It is possible to associate to block $k$ a new random variable $X_k^{(N)} = \frac{1}{N^3} \sum_{i \in k} X_i^{(1)}$. The probability distribution of the random variable $X_k^{(N)}$ is centered on the average chemical composition of the material $\delta$. Its characteristics depend on the possible correlations between the $X_i^{(1)}$ inside the block $k$.

As above, we define the following incomplete partition function

\begin{eqnarray}
\nonumber
\tilde{Z}^{(N)}(\{ {\bf u_k^{(N)} } \};\{ x_i^{(1)} \}) = 
C \int ... \int 
\{ \prod_{blocks~k} \delta(\sum_{i \in k} {\bf u_i^{(1)} } -N^3 {\bf u_k^{(N)} })  \}
e^{- \beta H^{(1)}(\{ {\bf u_i^{(1)}} \};\{ x_i^{(1)} \})}  \{ \prod_i d {\bf u_i^{(1)} }  \},
\end{eqnarray}

and the associated coarse-grained free energy, $H^{(N)}(\{ {\bf u_k^{(N)} } \} ;\{ x_i^{(1)} \})$, as 

\begin{equation}
H^{(N)}(\{ {\bf u_k^{(N)} } \};\{ x_i^{(1)} \}) = - k_B T ln \tilde{Z}^{(N)}(\{ {\bf u_k^{(N)} } \};\{ x_i^{(1)} \} ) 
\end{equation}

The main difference with the previous section is that the different blocks are not equivalent, owing to the chemical composition that might not be the same from a block to another.

Eq.~\ref{meanforce} extends straightforwardly to the present case:

\begin{equation}
\frac{ \partial H^{(N)} }{\partial {\bf u_n^{(N)} }}  (\{ {\bf u_k^{(N)} } \};\{ x_i^{(1)} \}) = 
- < \sum_{i \in n} {\bf f_i } > (\{ {\bf u_k^{(N)} } \};\{ x_i^{(1)} \})
\end{equation}

The derivative with respect to ${\bf u_n^{(N)} }$ takes the following form:

\begin{eqnarray}
\nonumber
\frac{ \partial H^{(N)} }{\partial {\bf u_n^{(N)} } }  (\{ {\bf u_k^{(N)} } \};\{ x_i^{(1)} \}) =  
\nonumber
\sum_{i \in n} 
<\frac{\partial H_{loc}^{(1)}}{\partial {\bf u} } ({\bf u_i^{(1)} };x_i^{(1)})> (\{ {\bf u_k^{(N)} } \};\{ x_i^{(1)} \})  \\
\nonumber
+\sum_{i \in n} 
\sum_{j \in n \atop (j \ne i)} \bar{\bar{C}}^{(1)}(i,j). <{\bf u_j^{(1)} }> (\{ {\bf u_k^{(N)} } \}; \{ x_i^{(1)} \})  
+ \sum_{n' \atop (n'\ne n)} \{
\sum_{i \in n \atop j \in n'} \bar{\bar{C}}^{(1)}(i,j) . <{\bf u_j^{(1)} }> (\{ {\bf u_k^{(N)} } \};\{ x_i^{(1)} \})
\}    
\end{eqnarray}

Up to this stage, the treatment is exact, and to go further, we make the following approximations :

\begin{enumerate}

\item Local Homogeneity : 

$\forall j \in n$, $<{\bf u_j^{(1)} }> (\{{\bf u_k^{(N)} } \};\{ x_i^{(1)} \}) = {\bf u_n^{(N)} }$,

\item We make the supplementary approximation that the average under fixed $\{{\bf u_k^{(N)} } \}$ 
of any function in the $u_{i,\alpha}^{(1)}$, of the form  $< {u_{i,\alpha}^{(1)}}^2 u_{i,\beta}^{(1)} > (\{ {\bf u_k^{(N)} } \};\{ x_i^{(1)} \})$, etc ... 
is a function of the local block variable to which $i$ belongs only, namely ${\bf u_n^{(N)} }$ (not of that of the neighboring blocks), and {\it of the local average chemical composition of the block $x_n^{(N)}$}:

$\forall i \in n,
< {u_{i,\alpha}^{(1)}}^2 u_{i,\beta}^{(1)} > (\{ {\bf u_k^{(N)} } \};\{ x_i^{(1)} \}) = 
< {u_{i,\alpha}^{(1)}}^2 u_{i,\beta}^{(1)} > ( {\bf u_n^{(N)} };x_n^{(N)} )$

It also depends naturally on the temperature.
These functions appear in the mean force when deriving the local part of $H^{(1)}$.

\end{enumerate}

These assumptions imply, here again, a block size lower than the correlation length~\cite{geneste2010,troster2005}, at least for the blocks the richest (or the poorest) in the polar species. 
Indeed, under fixed block variables, the different chemical species (Zr, Ti) have not the same behavior, which roughly results in two mean values of the local modes in the block. Thus assumption (1) can be satisfied only in the limit $x \rightarrow$ 0 or 1. However, in the following, only Ti-rich blocks with composition $x$=0.75, 0.875 and 1 will be considered (also for practical reasons related to the Kinetic Monte Carlo, see hereafter).

We continue deriving the formalism assuming these two hypothesis as fulfilled.
Under such approximations, the derivative depends only on the $\{  x_k^{(N)} \}$, and takes the following simple form,

\begin{eqnarray}
\nonumber
\frac{ \partial H^{(N)} }{\partial {\bf u_n^{(N)} }}  (\{ {\bf u_k^{(N)} } \};\{ x_k^{(N)} \}) \approx 
\sum_{i \in n} 
\{
<\frac{\partial H_{loc}^{(1)}}{\partial {\bf u }} ({\bf u_i^{(1)} };x_i^{(1)})> ({\bf u_n^{(N)} };x_n^{(N)}) 
+ \sum_{j \in n \atop (j \ne i)} \bar{\bar{C}}^{(1)}(i,j). {\bf u_n^{(N)} }
\} 
 + \sum_{n' \atop (n' \ne n)} \{
\sum_{i \in n \atop j \in n'} \bar{\bar{C}}^{(1)}(i,j) \} .  {\bf u_{n'}^{(N)} } 
\end{eqnarray}

After integration, we obtain the following form for the coarse-grained free energy:

\begin{eqnarray}
\label{hn2}
\nonumber
H^{(N)}  (\{ {\bf u_k^{(N)} } \};\{ x_k^{(N)} \}) \approx
\sum_{n} H_{loc}^{(N)} ({\bf u_n^{(N)} } ; x_n^{(N)}) +  
\frac{1}{2} \sum_{n,n' \atop (n' \ne n)} 
{\bf u_{n}^{(N)} }.  \bar{\bar{C}}^{(N)} (n,n') . {\bf u_{n'}^{(N)} } 
\end{eqnarray}

with the same notations as previously:

\begin{eqnarray}
\nonumber
\bar{\bar{C}}^{(N)} (n,n') = \sum_{i \in n} \sum_{j \in n'}  \bar{\bar{C}}^{(1)}(i,j) =
\sum_{i \in n} \sum_{j \in n'}  \bar{\bar{C}}^{(1)}_{SR}(i,j) + 
\sum_{i \in n} \sum_{j \in n'}  \bar{\bar{C}}^{(1)}_{LR}(i,j) =
\bar{\bar{C}}^{(N)}_{SR} (n,n') + \bar{\bar{C}}^{(N)}_{LR} (n,n')
\end{eqnarray}

The local part of $H^{(N)}$ (local free energy $H_{loc}^{(N)}$) is defined by its derivatives:

\begin{eqnarray}
\nonumber
\frac{\partial H_{loc}^{(N)}}{\partial {\bf u }} ({\bf u_n^{(N)} };x_n^{(N)})=
\sum_{i \in n} 
\{
<\frac{\partial H_{loc}^{(1)}}{\partial {\bf u }} ({\bf u_i^{(1)} };x_i^{(1)})> ({\bf u_n^{(N)} };x_n^{(N)}) 
+ \sum_{j \in n \atop (j \ne i)} \bar{\bar{C}}^{(1)}(i,j). {\bf u_n^{(N)} }
\} 
\end{eqnarray}

Finally, our assumptions allow $H^{(N)}$ to have the same functional form as $H^{(1)}$:

\begin{eqnarray}
\nonumber
H^{(N)}  (\{ {\bf u_k^{(N)} } \};\{ x_k^{(N)} \}) \approx
\sum_{n} H_{loc}^{(N)} ({\bf u_n^{(N)} } ; x_n^{(N)}) +  
\frac{1}{2} \sum_{n,n' \atop (n' \ne n)} 
{\bf u_n^{(N)} }.  \bar{\bar{C}}^{(N)}_{SR} (n,n') . {\bf u_{n'}^{(N)} } +
\frac{1}{2} \sum_{n,n' \atop (n' \ne n)} 
{\bf u_n^{(N)} }.  \bar{\bar{C}}^{(N)}_{LR} (n,n') . {\bf u_{n'}^{(N)} }
\end{eqnarray}

In the following, we focus on the relaxor case only, and consider all the above-mentioned assumptions as fulfilled.

\subsection{Block size}

The main condition that underlies the construction of the previous approximate form is that {\it the block size must be lower than the correlation length:} $Na_0 < \xi(T)$. It ensures that the local modes are correlated all over the block, so that the notion of "collective motion" makes sense. Each block contains thus, at most, on average, one polar nanoregion.

At given temperature, several values of $N$ are {\it a priori} possible. According to the value of $N a_0$ relative to $\xi(T)$, the coarse-grained hamiltonian may write differently:

\begin{figure}[htbp]
    {\par\centering
    {\scalebox{0.55}{\includegraphics{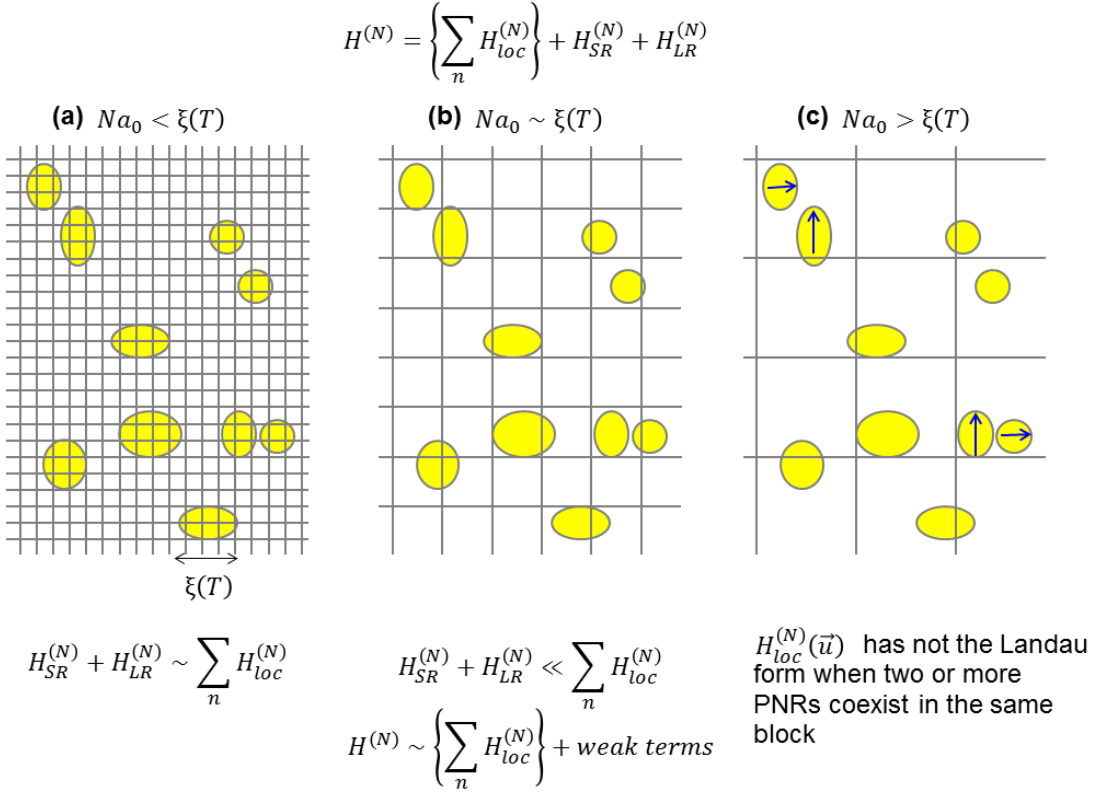}}}
    \par}
     \caption{{\small Evolution of the coarse-grained Hamiltonian $H^{(N)}$ at given temperature $T$ with the block size $N a_0$, i.e. for different RG transformations. The yellow areas feature the PNRs (in BZT, regions where Ti is more abundant than Zr).
     (a) $N a_0 < \xi$: the PNRs can extend over several blocks, making inter-block interactions very strong;
     (b) $N a_0 \sim \xi$: there is at most one PNR per block; such configuration minimizes the inter-block interactions while keeping for the local potential a smoothly-evolving Landau-like form;
     (c) $N a_0 > \xi$: several PNRs can coexist in a single block, and the local potential has not the Landau form any more.}}
    \label{xi}
\end{figure}

\begin{itemize}
 \item For $Na_0 < \xi$: one PNR can extend over several blocks. The blocks should be uniformly polarized but the inter-block interactions should be strong, with probably the same order of magnitude as the local part: $H_{SR}^{(N)}+H_{LR}^{(N)} \sim  \sum_n H_{loc}^{(N)}$. The local free energy $H_{loc}^{(N)}$ should thus be a smooth function of ${\bf u}$ and should be well approximated by a Landau expansion.
 
 \item For $Na_0 \sim \xi$: there is on average at most one PNR per block. The inter-block interactions are thus probably weaker, and $H_{loc}^{(N)}$ should be a smooth Landau-like function of ${\bf u}$. If these interactions are weak enough, the coarse-grained Hamiltonian has possibly the very simple form: $H^{(N)} \approx   \sum_n H_{loc}^{(N)} ({\bf u_n^{(N)} }; x_n^{(N)})$ + weaker terms. The blocks become weakly interacting.
  
 \item For $Na_0 > \xi$: some blocks might contain several PNRs. 
In that case, $H_{loc}^{(N)}$ is not a smoothly-evolving function of ${\bf u}$ any more, because a given value of ${\bf u}$ can hide several states of the PNRs inside the block.
\end{itemize}

Thus, choosing $N a_0 \sim \xi(T)$ should provide a quite simple form for the coarse-grained Hamiltonian. This form is the basis of our modeling of the time-dependent dielectric response in BZT (see last section):

\begin{eqnarray}
H^{(N)}  (\{ {\bf u_k^{(N)} } \};\{ x_k^{(N)} \}) \approx \sum_{n} H_{loc}^{(N)} ({\bf u_n^{(N)} }; x_n^{(N)})  
\label{glass}
 + weak~terms.
\end{eqnarray}

The possible effect of the block size $N a_0$ on the coarse-grained Hamiltonian is illustrated on Fig.~\ref{xi}.

\subsection{Landau expansion of the local free energy $H_{loc}^{(N)}({\bf u } ;x)$}

For block sizes lower than $\xi$(T), the local modes should be sufficiently correlated within the blocks, so that the previous model can work. One expects a smooth and continuous evolution of the local part of $H^{(N)}$ with ${\bf u}$. Thus it can be expanded in power series of the local order parameters ${\bf u}$, truncated to fourth order as explained above. The coefficients of the expansion are dependent on the temperature and on the local chemical composition $x$. Moreover, we make the reasonable assumption that $H_{loc}^{(N)}$ possesses the symmetry of the parent cubic lattice:

\begin{eqnarray}
\nonumber
H_{loc}^{(N)}({\bf u } ;x) - H_{loc}^{(N)}({\bf 0 } ;x) = 
\nonumber
\kappa_{2}^{(N)} (T;x) (u_X^2 + u_Y^2 + u_Z^2) +
\alpha^{(N)}(T;x) || {\bf u} ||^4 
\label{landau}
+ \gamma^{(N)}(T;x) (u_X^2 u_Y^2 + u_X^2 u_Z^2 + u_Y^2 u_Z^2)
\end{eqnarray}

If we admit that the coefficients satisfy the basic assumptions of Landau theory of second-order phase transitions, namely $\kappa_{2}^{(N)} (T;x) = A(x) (T - T_c(x) )$, and $\alpha^{(N)}(T;x)=\alpha^{(N)}(x)$, $\gamma^{(N)}(T;x)=\gamma^{(N)}(x)$, we can define, locally in a block with chemical composition $x$, a "transition" temperature depending on $x$, $T_c(x)$. This way, $H^{(N)}$ has a functional form completely similar to that of $H^{(1)}$, even in the expansion of its local terms.

If the renormalized short-range and dipole-dipole interactions are weak with respect to the local free energy (e.g. $Na_0 \sim \xi$), the properties of the block are mainly controlled by the local part of $H^{(N)}$.
For $T \leq T_c(x)$, the blocks with chemical composition $x$ are polar, while for $T \geq T_c(x)$, they are non polar. This local polarity is not strongly modified by the inter-block interactions, that rather act as small perturbations: they modify slightly the position and depth of the local minima in each block, and slightly change the free energy barriers separating those minima. Note that $T_c(x)$ can be zero or negative if the block is intrinsically non-polar (for instance in BZT, Zr-rich blocks).

The picture emerging from Eq.~\ref{glass} is that of an inhomogeneous system, with local polar instabilities having amplitudes and local "transition" temperatures depending on the local chemical composition $x$. There is a distribution of local $T_c(x)$, the highest one, $T_{cm}$ corresponding in BZT to the blocks that are the richest in the chemical species that drives the FE instability~\cite{bzt}. The number of polar blocks increases as temperature decreases, and the polar instability becomes more and more pronounced in the polar blocks (the depth of the local free energy well increases). Also the free energy barriers that separate the minima of the local free energy in each block, and that control the dynamics of the PNRs, become larger and larger as T decreases.

\section{Computation of the coarse-grained Hamiltonian}

The basic ingredients necessary to perform numerical simulations (Monte Carlo) of relaxors in the coarse-grained hamiltonian framework are therefore, provided the previous approximations are valid : 

(i) the local free energy $H_{loc}^{(N)}({\bf u } ; x)$.

(ii) the matrix elements of the renormalized Short-Range interaction $\bar{\bar{C}}_{SR}^{(N)} (n,n')$;

(iii) the matrix elements of the renormalized Dipole-Dipole interaction $\bar{\bar{C}}_{LR}^{(N)} (n,n')$;

In the framework of these approximations, only the local part depends on the temperature.

\subsection{Local free energy $H_{loc}^{(N)}({\bf u } ; x)$: phenomenological form}

Ideally, $H_{loc}^{(N)}({\bf u } ; x)$ could be determined directly from microscopic simulations, using for instance the thermodynamic integrations methods that have been recently used with effective Hamiltonians~\cite{geneste2009,geneste2010,geneste2011,kumar2010}. 
This is not the method we have employed here. Instead, we have chosen to express the local free energy under a simplified phenomenological form. We start from the local free energy expressed for a Ti-rich block ($x$=1), written as an energy {\it per 5-atom cell}:

\begin{eqnarray}
H_{loc,Ti}^{(N)}({\bf u }) - H_{loc,Ti}^{(N)}({\bf 0 })=
a_1' (T-T_0) (u_X^2 + u_Y^2 + u_Z^2) + a_{11} (u_X^4 + u_Y^4 + u_Z^4),
\label{landau-ti}
\end{eqnarray}

that we extend to blocks of any composition $x$ according to

\begin{eqnarray}
\nonumber
H_{loc}^{(N)}({\bf u } ;x) - H_{loc}^{(N)}({\bf 0 } ;x) = x [H_{loc,Ti}^{(N)}(\frac{{\bf u }}{x}) - H_{loc,Ti}^{(N)}({\bf 0 })]
\label{landau}
\end{eqnarray}

The drawback of such form is that all the blocks have the same local transition temperature, $T_0$, whatever their chemical composition $x$. This is certainly not physical for the Zr-rich blocks ($x \rightarrow 0$), that probably never transit to a polar state. To circumvent this problem, in the Kinetic Monte Carlo simulation, only the Ti-rich blocks are considered ($x \geq 0.75$).

However, there is also a practical advantage to such form regarding the Kinetic Monte Carlo application: when the temperature approaches $T_c(x)$ from below, the free energy barriers for hopping tend to zero, making the corresponding events occurring very frequently, and parasiting the Kinetic Monte Carlo simulation. To be relevant, such simulation should retain only the events necessary to produce the dielectric response on the desired time-scale. In the previous form, since all the blocks have the same $T_c(x) = T_0$, this problem only appears at one temperature.

T$_0$ is identified to the temperature above which static PNRs disappear in Ref.~\onlinecite{bzt} (240 K).
To determine the two other coefficients, we use a 12 $\times$ 12 $\times$ 12 supercell with Ti and Zr randomly distributed (thus corresponding to BZT50), which is then divided in 2 $\times$ 2 $\times$ 2 blocks. One of this block is selected and made Ti-rich ($x$=1). Molecular Dynamics simulations are performed using the Hamiltonian $H^{(1)}$, under the constraint of fixed ${\bf  u_k^{(N)} } = {\bf 0}$ in all the blocks except the one selected ($n$). Fixing ${\bf u_k^{(N)} } = {\bf 0 }$ in all the blocks except $n$ reduces the coarse-grained Hamiltonian to the local free energy in block $n$: $H^{(N)} = H_{loc}^{(N)}({\bf u } ;x=1) = a_1' (T-T_0) (u_X^2 + u_Y^2 + u_Z^2) + a_{11} (u_X^4 + u_Y^4 + u_Z^4)$, where ${\bf u}$ is the value of the block variable in block $n$. The simulations are performed at very low temperature, so that the energy $E$ can be identified to the free energy:

\begin{equation}
\nonumber
E({\bf u }) - E({\bf 0 }) \approx - a_1' T_0 (u_X^2 + u_Y^2 + u_Z^2) + a_{11} (u_X^4 + u_Y^4 + u_Z^4)
\end{equation}

Several runs are performed by changing the Ti/Zr distribution. In each case, the mean local mode ${\bf u_0}$ of block $n$ is extracted, as well as the energy difference $E({\bf u_0 }) - E({\bf 0 })$. With these two data, it is possible to extract the $a_1'$ and $a_{11}$ coefficients: $a'_1$ = 0.053654 eV/(\AA$^2$.K), and $a_{11}$ = 420.8273 eV/\AA$^4$. Note that Eq.~\ref{landau-ti} enforces the minima of $H_{loc}^{(N)}$, i.e. the stable sites, to lie along the $<$111$>$ directions, which is qualitatively observed in BZT~\cite{bzt}.

\begin{figure}[htbp]
    {\par\centering
    {\scalebox{0.5}{\includegraphics{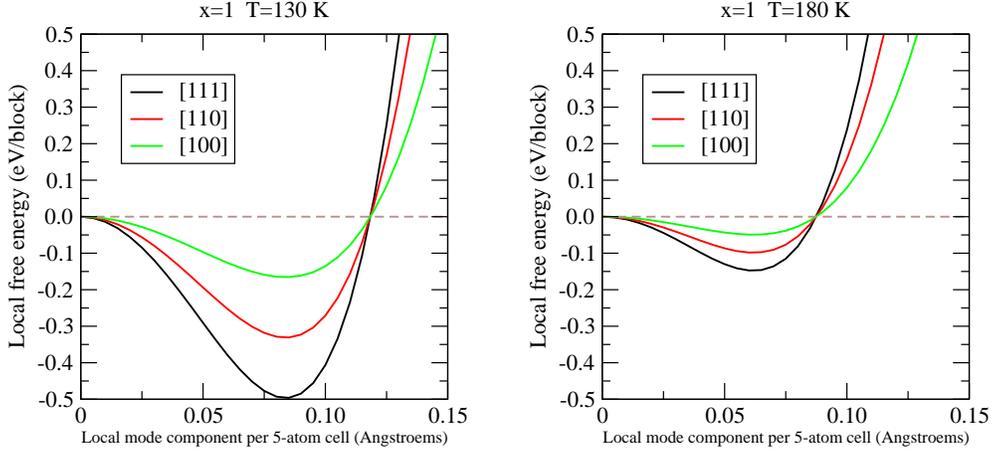}}}
    \par}
     \caption{{\small Local free energy per block (eV), for Ti-rich blocks ($x$=1) and for two temperatures (T=130 K and 180 K), as a function of mean local mode component along [111] ($u_X = u_Y = u_Z$), [110] ($u_X = u_Y$, $u_Z$=0) and [100] ($u_X$, $u_Y=u_Z$=0). The curves are plotted from Eq.~\ref{landau-ti} ($\times N^3$). The local mode components are expressed in lattice constants a$_0$. The free energy barrier for hopping is the difference between the minimum of red and black curves.}}
    \label{exemple-hloc}
\end{figure}

Fig.~\ref{exemple-hloc} shows how $H_{loc,Ti}^{(N=2)}({\bf u})$ (Ti-rich block) varies as a function of ${\bf u}$ along the [111], [110] and [100] directions for two temperatures, T = 130 K and 180 K. Note that the free energy plotted is for one block, i.e. the quantity of Eq.~\ref{landau-ti} with the coefficients given above has to be multiplied by $N^3$=8 here. It is important to understand that the free energy landscape that must be considered for estimating the transition rates corresponds to an energy {\it per block}, and not per unit cell, because such hopping corresponds to a collective motion of all the local modes inside the block.
The local free energy barrier for the thermally activated hopping of the block variable from one minimum to another is extracted as the difference of the minimum along [110] and that along [111].

Although very simple, this phenomenological form is able to capture the physics of the collective motions in the small 2 $\times$ 2 $\times$ 2 blocks.

\subsection{Renormalized Short-Range interaction}

Between blocks $n$ and $n'$, the renormalized Short-Range interaction matrix is $\bar{\bar{C}}_{SR}^{(N)}(n,n') = \sum_{i \in n} \sum_{j \in n'} \bar{\bar{C}}_{SR}^{(1)} (i,j)$. Its coefficients are obtained by summing those of the starting lattice by pair $(i,j), i \in n, j \in n'$. In this subsection, the SR matrix between 1st neighbors aligned along [100] is denoted as $\bar{\bar{C}}_{SR}^{(1)}(1,X)$, that between 2nd neighbors aligned along [110] is denoted as $\bar{\bar{C}}_{SR}^{(1)}(2,XY)$, and that between 3rd neighbors aligned along [111] is denoted as $\bar{\bar{C}}_{SR}^{(1)}(3,XYZ)$.

Since the short-range interaction extends up to third neighbors in the starting lattice, it is thus obvious that the renormalized short-range interaction also extends up to the third neighbor block only in the renormalized lattice. In the real lattice, the SR interaction is characterized by 7 coupling parameters $j_1^{(1)}$, $j_2^{(1)}$, $j_3^{(1)}$, $j_4^{(1)}$, $j_5^{(1)}$, $j_6^{(1)}$ and $j_7^{(1)}$, which are supposed to decay rapidly with distance:

\begin{equation}
 j_1^{(1)},~j_2^{(1)} >> j_3^{(1)},~j_4^{(1)},~j_5^{(1)} >> j_6^{(1)},~j_7^{(1)}
\end{equation}

The corresponding matrices write:

\begin{equation}
\bar{\bar{C}}_{SR}^{(1)}(1,X) =
\left( 
\begin{array}{ccc}
j_2^{(1)}  & 0    & 0 \\
0    & j_1^{(1)}  & 0 \\
0    &  0   &  j_1^{(1)}  \end{array} 
\right) 
\end{equation}

\begin{equation}
\bar{\bar{C}}_{SR}^{(1)}(2,XY) =
 \left( \begin{array}{ccc}
j_3^{(1)} &  j_5^{(1)} & 0  \\
j_5^{(1)} &  j_3^{(1)} & 0  \\
0      &  0  & j_4^{(1)} \end{array} \right)  
\end{equation}

\begin{equation}
\bar{\bar{C}}_{SR}^{(1)}(3,XYZ) =
 \left( \begin{array}{ccc}
j_6^{(1)} &  j_7^{(1)} & j_7^{(1)}  \\
j_7^{(1)} &  j_6^{(1)} & j_7^{(1)}  \\
j_7^{(1)} &  j_7^{(1)} & j_6^{(1)} \end{array} \right) 
\end{equation}

\begin{figure}[htbp]
    {\par\centering
    {\scalebox{0.3}{\includegraphics{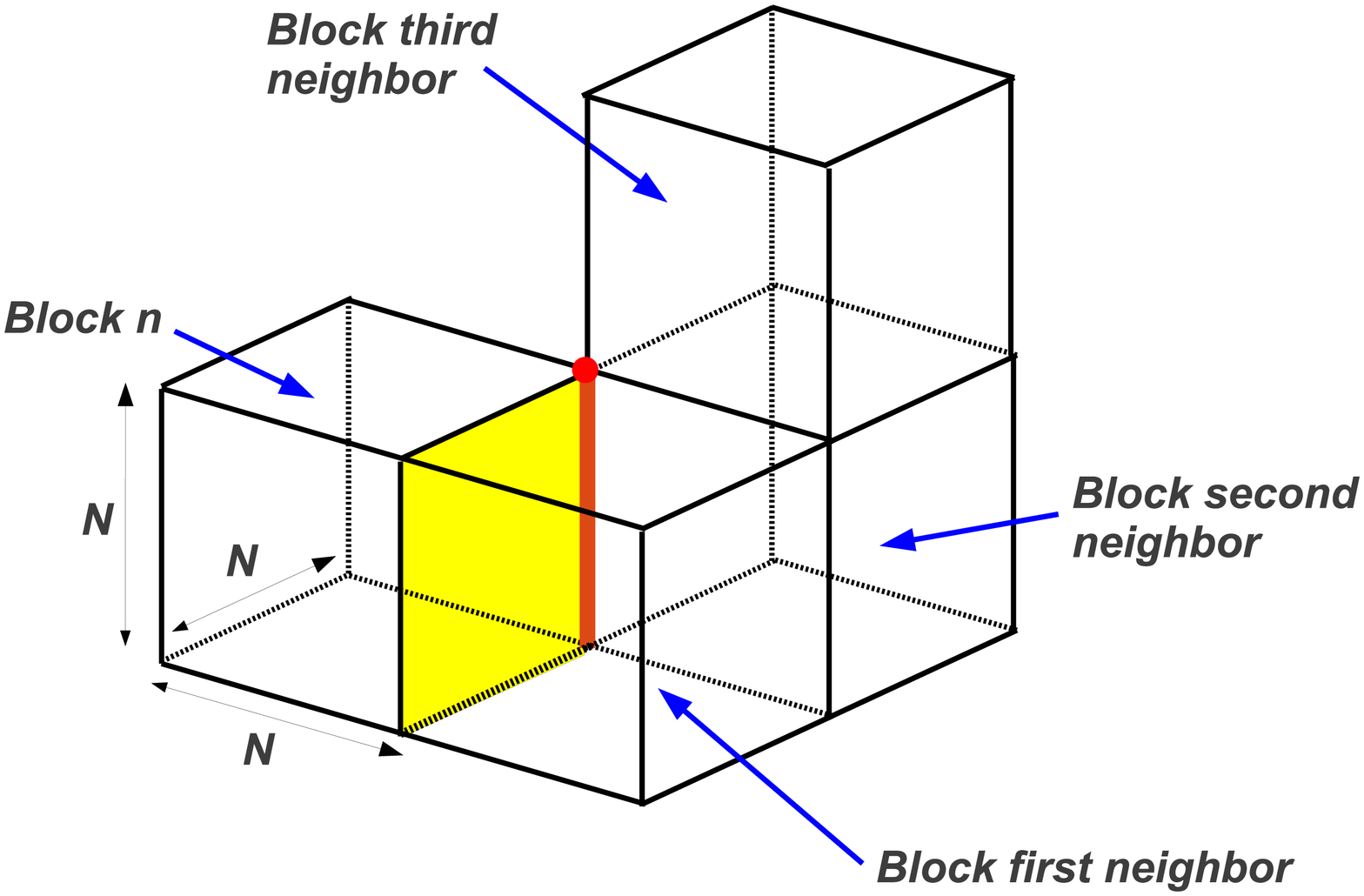}}}
    \par}
     \caption{{\small Calculation of the renormalized short-range interaction matrices between first, second and third neighbor blocks. Yellow area: the unit cells taken into account to compute $\bar{\bar{C}}_{SR}^{(N)}(1,X)$.
Orange line: the unit cells taken into account to compute $\bar{\bar{C}}_{SR}^{(N)}(2,XY)$.
Red point: the unit cell taken into account to compute $\bar{\bar{C}}_{SR}^{(N)}(3,XYZ)$.
     }}
    \label{short-range}
\end{figure}

To calculate $\bar{\bar{C}}_{SR}^{(N)}(n,n')$, we must count, for two blocks $n$ and $n'$ (1st, 2nd or 3rd neighbors) the number of couples of cells $(i,j)$, $i$ belonging to $n$, $j$ belonging to $n'$ which are 1st, 2nd or 3rd neighbors in the real lattice (Fig.~\ref{short-range}). 

The matrix of the renormalized SR interaction between two 1st neighbor blocks aligned along $X$ writes:

\begin{equation}
\bar{\bar{C}}_{SR}^{(N)}(1,X) = 
 \left( \begin{array}{ccc}
N^2 j_2^{(1)} + 4N(N-1)j_3^{(1)} + \atop (2N-2)^2 j_6^{(1)} & 0 & 0 \\
0           &  N^2 j_1^{(1)} + 2N(N-1)(j_3^{(1)}+j_4^{(1)}) + \atop (2N-2)^2 j_6^{(1)}  & 0 \\
0           &  0 & N^2 j_1^{(1)} + 2N(N-1)(j_3^{(1)}+j_4^{(1)}) + \atop (2N-2)^2 j_6^{(1)} \end{array} \right) 
\end{equation}

Considering the rapid decay of the SR interaction in the real lattice, we can roughly approximate this matrix by

\begin{equation}
\bar{\bar{C}}_{SR}^{(N)}(1,X) \approx
 \left( \begin{array}{ccc}
N^2 j_2^{(1)}  & 0 & 0 \\
0            &  N^2 j_1^{(1)}   & 0 \\
0            &  0  & N^2 j_1^{(1)}  \end{array} \right) 
\end{equation}

The matrix of the renormalized SR interaction between two 2nd neighbor blocks aligned along $X+Y$ writes:

\begin{equation}
\bar{\bar{C}}_{SR}^{(N)}(2,XY) = 
 \left( \begin{array}{ccc}
Nj_3^{(1)}+2(N-1)j_6^{(1)} &  Nj_5^{(1)} +2(N-1)j_7^{(1)} & 0  \\
Nj_5^{(1)} +2(N-1)j_7^{(1)} &  Nj_3^{(1)}+2(N-1)j_6^{(1)} & 0  \\
0     &  0        & Nj_4^{(1)}+2(N-1)j_6^{(1)} \end{array} \right) 
\end{equation}

Considering the rapid decay of the SR interaction in the real lattice, we can roughly approximate this matrix by

\begin{equation}
\bar{\bar{C}}_{SR}^{(N)}(2,XY) \approx
 \left( \begin{array}{ccc}
Nj_3^{(1)} &  Nj_5^{(1)} & 0  \\
Nj_5^{(1)} &  Nj_3^{(1)} & 0  \\
0      &  0      & Nj_4^{(1)} \end{array} \right) 
\end{equation}

The matrix of the renormalized SR interaction between two 3rd neighbor blocks aligned along $X+Y+Z$ writes:

\begin{equation}
\bar{\bar{C}}_{SR}^{(N)}(3,XYZ) = \bar{\bar{C}}_{SR}^{(1)}(3,XYZ) =
 \left( \begin{array}{ccc}
j_6^{(1)} &  j_7^{(1)} & j_7^{(1)}  \\
j_7^{(1)} &  j_6^{(1)} & j_7^{(1)}  \\
j_7^{(1)} &  j_7^{(1)} & j_6^{(1)} \end{array} \right) 
\end{equation}

Identification of the approximate forms of these renormalized SR matrices with the starting ones provides the renormalization scheme given in Tab.~\ref{renorm}.

Note that, as expected, the renormalized matrices have the same form as the starting one, i.e. they possess the cubic symmetry of the parent lattice.

\subsection{Renormalized Dipole-Dipole interaction}

Between blocks $n$ and $n'$, the renormalized Dipole-Dipole interaction matrix is $\bar{\bar{C}}_{LR}^{(N)}(n,n') = \sum_{i \in n} \sum_{j \in n'} \bar{\bar{C}}_{LR}^{(1)} (i,j)$. The elements of this matrix are:

\begin{equation}
\nonumber 
\bar{\bar{C}}_{LR,\alpha,\beta}^{(1)} (i,j) = 
\frac{{Z^{*}}^2}{\epsilon_{\infty}}
\frac{\delta_{\alpha \beta} - 3 {\hat{r}}_{ij,\alpha} {\hat{r}}_{ij,\beta} }{R_{ij}^3},
\end{equation}

where $R_{ij}$ is the distance that separates cell $i$ from cell $j$, and $\hat{{r}}_{ij} = {\bf R_{ij} } / R_{ij}$.

Here again, we must sum these terms over pairs $(i,j), i \in n, j \in n'$.

One approximation is to consider that the interaction between the macro-dipoles associated to blocks $n$ and $n'$ are the main contribution to this matrix. The macro-dipole associate to block $n$ (resp. $n'$) is $N^3 {Z^{*}} {\bf  u_n^{(N)} }$ (resp.  $N^3 {Z^{*}} {\bf  u_{n'}^{(N)} } $) and their interaction, namely,
$\approx  u_{n,\alpha}^{(N)} . \bar{\bar{C}}_{LR,\alpha,\beta}^{(N)} (n,n'). u_{n',\beta}^{(N)}$, can be approximated by:

\begin{equation}
\bar{\bar{C}}_{LR,\alpha,\beta}^{(N)} (n,n') \approx
\frac{({Z^{*}N^{3}})^2}{\epsilon_{\infty}}
\frac{\delta_{\alpha \beta} - 3 {\hat{r}}_{nn',\alpha} {\hat{r}}_{nn',\beta} }{R_{nn'}^3},
\end{equation}

In summary, Tab.~\ref{renorm} lists the coefficients of the starting Hamiltonian $H^{(1)}$ and their renormalized equivalent in $H^{(N)}$. The values for $H^{(N)}$ can directly replace those of $H^{(1)}$ in any simulation (e.g. Monte Carlo, Molecular Dynamics).

\begin{table*}
\caption{\label{renorm}\small Parameters of the effective Hamiltonian $H^{(1)}$, and corresponding parameters (coupling constants) of the renormalized Hamiltonian $H^{(N)}$. $x^{(1)}$ characterizes the local disorder in the real sublattice (it takes two values 0 and 1), while $x$ characterizes the local disorder in the renormalized lattice, thus at the scale of the block, 0$\leq x \leq$ 1.}
\begin{ruledtabular}
\begin{tabular}{lcc}
Hamiltonian   &  $H^{(1)}$  &  $H^{(N)}$  \\
\hline
Lattice constant & a$_0$  &  $N$ a$_0$  \\
\hline
Local potential & & \\
Quadratic coefficients   &   $\kappa_{2}^{(1)}(A)$ $~~~x^{(1)}$=0   &   $\kappa_{2}^{(N)} (T;x)$ \\
                        &    $\kappa_{2}^{(1)}(B)$ $~~~x^{(1)}$=1   &    \\
Quartic coefficients   &   $\alpha^{(1)}(A)$ $~~~x^{(1)}$=0   &   $\alpha^{(N)}(T;x)$    \\
        &   $\alpha^{(1)}(B)$ $~~~x^{(1)}$=1   &     \\
   &    $\gamma^{(1)}(A)$ $~~~x^{(1)}$=0   &   $\gamma^{(N)}(T;x)$   \\
   &    $\gamma^{(1)}(B)$ $~~~x^{(1)}$=1   &     \\   
\hline
Short-Range Interaction  & & \\
First neighbors   & $j_1^{(1)}$ &   $N^2 j_1^{(1)} + 2N(N-1)(j_3^{(1)}+j_4^{(1)}) + (2N-2)^2 j_6^{(1)} \approx  N^2 j_1^{(1)}$ \\
   & $j_2^{(1)}$ & $N^2 j_2^{(1)} + 4N(N-1)j_3^{(1)} + (2N-2)^2 j_6^{(1)} \approx  N^2 j_2^{(1)}$ \\
Second neighors   &       $j_3^{(1)}$ & $Nj_3^{(1)}+2(N-1)j_6^{(1)} \approx N j_3^{(1)}$   \\
   & $j_4^{(1)}$ & $Nj_4^{(1)}+2(N-1)j_6^{(1)} \approx N j_4^{(1)}$  \\
   & $j_5^{(1)}$ & $N j_5^{(1)} +2(N-1)j_7^{(1)} \approx N j_5^{(1)}$ \\
Third neighbors   & $j_6^{(1)}$ & $j_6^{(1)}$  \\
   & $j_7^{(1)}$ & $j_7^{(1)}$   \\
\hline
Dipole-Dipole Interaction  & & \\
Effective charge   & $Z^{*}$   & $N^{3} Z^{*}$ (= $Z^{*}_b$) \\
\end{tabular}
\end{ruledtabular}
\end{table*}

\section{Kinetics}

In this part, we now explain how to make the connection between the previous thermodynamic aspects and the dynamics of the system. 
Basically, such connection can be made using Transition State Theory (TST, see hereafter). In TST, the transition rate corresponding to a thermally activated process (particle hopping over a barrier) depends on a free energy barrier, and from a prefactor that involves the mass of the particle.

Thus, one needs first to associate a mass $M$ to the block variables. This mass $M$ is expected, if the block is large enough, to be much larger than the mass of the local modes, $M >> m$, which is of the order of the atomic masses. This is equivalent to assume that the block variables are slow degrees of freedom with respect to all the microscopic ones, which have been integrated out in the building of $H^{(N)}$. Reasonable choice for this mass should allow modeling the dynamics of the block variables. This mass $M$ is expected to be involved in Newton-like equations of motion of the block variables, subject to forces deriving from the coarse-grained Hamiltonian:

\begin{equation}
M \frac{d^2 {\bf u_n^{(N)} } }{dt^2} = - \frac{\partial H^{(N)}}{\partial {\bf u_n^{(N)} } } (\{  {\bf u_k^{(N)} } \})
\end{equation}

\subsection{Mass of the local order parameters}

The goal of this section is to obtain an {\it order of magnitude} of this mass, that will be used hereafter to estimate the transition rate prefactors. For simplicity, we assume therefore all the local modes to have the same mass $m$, although this is not true in the case of BZT ($m_{Zr} \ne  m_{Ti}$). Note, however, that the blocks considered in the KMC are only the Ti-richest ones ($x \geq$ 0.75).

We start from the Newton equations of motion for each local mode

\begin{equation}
\nonumber
m \frac{d^2 {\bf u_i^{(1)} } }{dt^2} = {\bf f_i },
\end{equation}

that we sum over the cells belonging to a given block $n$:

\begin{equation}
\nonumber
m  \frac{d^2}{dt^2} \sum_{i \in n} {\bf u_i^{(1)} } = \sum_{i \in n} {\bf f_i }
\end{equation}

Then we introduce the block variable ${\bf u_n^{(N)} } = \frac{1}{N^3} \sum_{i \in n} {\bf u_i^{(1)} } $, and obtain

\begin{equation}
\nonumber
m N^3  \frac{d^2 {\bf u_n^{(N)} } }{dt^2}  = \sum_{i \in n} {\bf f_i }
\end{equation}

At this stage, we can assume that we are mainly interested in the "slow" part of the dynamics of ${\bf  u_n^{(N)} }$, i.e. the component of its motion which is much slower than the atomic motions. We postulate there is a time scale $\Delta t$ which is (i) much longer than the atomic characteristic times ($\sim$ 10$^{-13}$ s) and (ii) much smaller than the typical time scales of the phenomena involved in the radiofrequency dielectric processes of relaxors (from $\sim$ 10$^{-3}$ s to $\sim$ 10$^{-8}$ s). Typically, $\Delta t$ can be chosen $\sim$ 10$^{-9}$ s. 

We proceed to a time-averaging of the previous equations over a few $\Delta t$,

\begin{equation}
\label{condtime}
m N^3 < \frac{d^2 {\bf u_n^{(N)} } }{dt^2} >_{\Delta t} = <\sum_{i \in n} {\bf  f_i } >_{\Delta t}
\end{equation}

${\bf  u_n^{(N)} }(t)$ can be written as the sum of a slowly varying part $ {\bf U_n^{(N)} } (t)$ (considered as constant over the time scale $\Delta t$, as well as its time derivatives) and a rapidly varying part ${\bf \delta u_n^{(N)} } (t)$ reflecting the microscopic motions (its average and the average of its derivative are zero over a few $\Delta t$)~\cite{kramers}. Derivating twice and time-averaging over $\Delta t$ allows to write

\begin{equation}
\nonumber
< \frac{d^2 {\bf u_n^{(N)} } }{dt^2} >_{\Delta t} = \frac{d^2 {\bf U_n^{(N)} } }{dt^2}
\end{equation}

It is possible to precise the time-average of the force in Eq.~\ref{condtime}, because

(i) all the microscopic degrees of freedom, except the slow $\{  {\bf U_k^{(N)} } \}$, are supposed to be thermalized over a few $\Delta t$ ($\Delta t$ is much larger than the phonon relaxation times);

(ii) such time-average corresponds to a thermal average under fixed $\{  {\bf U_k^{(N)} } \}$ and fixed $\{  {\bf \dot{U}_k^{(N)} } \}$, and is in fact equal to the conditional average at fixed $\{  {\bf U_k^{(N)} } \}$: this is because the fixed $\{  {\bf U_k^{(N)} } \}$ are linear combinations of the configuration variables~\cite{sprik1998}.

Thus, the time-average of the force can be identified to a conditional average at fixed $\{  {\bf U_k^{(N)} } \}$, as defined at the beginning of this document, and is equal to minus the derivative of the coarse-grained Hamiltonian:

\begin{equation}
\label{newton}
m N^3  \frac{d^2 {\bf U_n^{(N)} } }{dt^2}  = <\sum_{i \in n} {\bf f_i } > (\{  {\bf U_k^{(N)} } \}) =
- \frac{\partial H^{(N)}}{\partial {\bf u_n^{(N)} } } (\{  {\bf U_k^{(N)} } \})
\end{equation}

The mass associated with the $\Delta t$-averaged local order parameters $\{ {\bf U_k^{(N)} }  \}$ is thus simply the total mass of (the local modes of) the block, $M= m \times N^3 $. Note also that $ {\bf u_n^{(N)} } $ is the mass center of the local modes of the block. For time scales $> \Delta t$, the $\{ {\bf U_k^{(N)} }  \}$ can be seen as evolving over the energy landscape described by the coarse-grained Hamiltonian:

\begin{equation}
\label{newton2}
M  \frac{d^2 {\bf U_n^{(N)} } }{dt^2}  = - \frac{\partial H^{(N)}}{\partial {\bf u_n^{(N)} } } (\{  {\bf U_k^{(N)} } \})
\end{equation}

Finally we point out that these equations of motion (Eqs.~\ref{newton}) cannot be used as such in a Molecular Dynamics simulation to obtain kinetic data, since a fundamental ingredient is missing: the thermal agitation has been completely removed by the time-averaging process. These equations are thus not able to simulate the occurrence of the rare events, which are thermally activated. 
One possibility would be to restaure the brownian motion of $ {\bf u_k^{(N)} } (t) = {\bf U_k^{(N)} } (t) + {\bf \delta u_k^{(N)} }(t) $ by means of a Langevin equation~\cite{kramers}.

\subsection{Transition rate of the local order parameter}

Simulation of the dynamical properties on long time scales is easier by using the Kinetic Monte Carlo method. For that, we need to estimate the transition rates of the block variables to jump from a local minimum of $H^{(N)}$ to another. We use the same arguments as the ones that fund the Transition State Theory (TST) in Chemistry. TST is precisely based on a set of hypothesis that allow to compute a transition rate from thermodynamic ingredients. These hypothesis are the following:

\begin{enumerate}

\item Motions are classical (no quantum correction applied)\footnote{as a consequence, all the hopping events considered are produced by thermal agitation (over-barrier motions), not by tunneling.};

\item The process is controlled by a parameter $\lambda$ (reaction coordinate), that is used to define a path between the initial state (i) and the final state (f). Also, an incomplete free energy is defined as a function of $\lambda$;

\item Along the path, there is a state (c) of highest incomplete free energy (transition state). The probability to reach the transition state is governed by the canonical distribution, i.e. the ratio of the equilibrium probabilities between (c) and (i) is  $e^{ - \Delta \tilde{F} (T)/ k_B T}$, with $\Delta \tilde{F} (T)$ the difference of incomplete free energy between (c) and (i) ("Local equilibrium" hypothesis);

\item Once in the transition state, the system systematically undergoes the process and falls into (f) ("no recrossing" hypothesis).

\end{enumerate}

Under such hypothesis, the transition rate from (i) to (f) is given by 
\begin{equation}
\label{rate}
r_{if}^{TST} = \sqrt{\frac{k_B T}{2 \pi M}}.P(\lambda_i) e^{- \Delta \tilde{F} (T) / k_B T},
\end{equation}

where $M$ is the mass associated with the reaction coordinate and $P(\lambda_i)$ the density of probability of the reaction coordinate in the initial state. We keep the previous hypothesis excepting (2), since in our case the problem is multidimensional: the local order parameters $\{ {\bf  u_n^{(N)} }  \}$ play the role of a set of reaction coordinates, with masses $M = m N^3$, and the incomplete free energy to be considered is precisely the coarse-grained hamiltonian $H^{(N)}$. 
The $\{ {\bf  u_n^{(N)} }  \}$ can be seen as heavy particles trapped during long time scales in the local minima of $H^{(N)}$, and jumping from time to time into the next minimum upon thermal agitation.

In a given block $n$ with chemical composition $x$, the local free energy $H_{loc}({\bf  u };x)$ usually exhibits several minima, which constitute the most probable states of ${\bf  u_n^{(N)} }$. There exists a minimum Free energy path (MFEP) separating each couple (A,B) of these minima, and this MFEP is associated to a transition state, which is the point of higher free energy along the path. Let us note the free energy barrier from A to B as $\Delta F_{AB}(T)$. Since the prefactor of the transition rate is typically proportional to $M^{-1/2}$, we can estimate the transition rate from A to B by 

\begin{equation}
\label{rate}
r_{AB} = r'_0 N^{-3/2} e^{- \Delta F_{AB}(T) / k_B T},
\end{equation}

with $r'_0$ a typical atomic attempt frequency.
In BZT, the blocks are taken to be 2 $\times$ 2 $\times$ 2 (8 unit cells). The prefactor of the transition rates, $r_0$, in the Kinetic Monte Carlo simulation is set at $r_0 = r'_0 N^{-3/2}$ = 1.0 $\times$ 10$^{12}$ Hz.

\subsection{Kinetic Monte Carlo}

Once the method to calculate the transition rates of the block variables has been established, it is possible to follow their dynamics by performing a Kinetic Monte Carlo (KMC) simulation. This method provides a numerical solution of the master equation by constructing a trajectory in which the elementary events are chosen and carried out using an algorithm based on random numbers. For that, the configuration space is considerably reduced: a microscopic state consists in providing the state of each of the $\{ {\bf  u_k^{(N)} } \}$ on a rigid lattice, i.e. each block variable has only a finite set of possible values corresponding to one of the 8 minima of the local free energy.

The KMC method is based on the so-called "residence-time" algorithm, described hereafter. 
At each step:

\begin{enumerate}
\item The complete list of the possible elementary events is made. Here an elementary event consists in the motion of one given block variable, from a local minimum of $H^{(N)}$ to a next one. Let us label these events by $p$ ($p=1...p_{max}$). Each event is characterized by a MFEP and a free energy barrier $\Delta F_{p}(T)$.

\item For each event $p$, the corresponding transition rate $r_p$ is computed according to $r_p = r_0 e^{- \Delta F_{p}(T) / k_B T}$.

\item A first random number $y_1$ is drawn between 0 and 1 with uniform probability.

\item One event is selected in the list, with a probability proportional to its transition rate. Practically, this is done by computing the cumulative quantities $R_p = \sum_{k=1}^p  r_k$ for all $p$, and choosing the event number $m$ that satisfies
$R_m  \geq y_1 R  \geq  R_{m-1}$, with $R = R_{pmax} = \sum_{k=1}^{p_{max}} r_k$.

\item This event is carried out, and the microscopic configuration accordingly updated.

\item A second random number $y_2$ is drawn between 0 and 1 with uniform probability, and the clock is updated by $\Delta t = - \frac{ln(y_2)}{R}$.

\end{enumerate}

Application of this KMC algorithm to the present system implies that two assumptions are valid: (i) the time increment $\Delta t$ is small compared to the period of the external field, and (ii) the supercell is large enough so that long-range interactions with periodic images do not modify significantly the energy barriers. See Ref.~\onlinecite{kmcktl2011} for details.

\subsection{Construction of the supercell and KMC simulations}

\subsubsection{Supercell}
A supercell of blocks is generated as follows: we start from a supercell containing 24 $\times$ 24 $\times$ 24 unit cells. To each of these unit cells is affected a chemical composition (Zr or Ti) randomly, with equiprobability. Then we gather them in 2 $\times$ 2 $\times$ 2 blocks, and the chemical composition of each block is computed (it can take the 9 values between 0 and 1 by step of 0.125). The KMC simulations are then conducted on this supercell of 12 $\times$ 12 $\times$ 12 blocks, each having its own local composition.

\subsubsection{Parameters of the KMC simulations}
Our KMC code is similar to the one used in Ref.~\onlinecite{kmcktl2011}.

Once the supercell constructed, the blocks are divided in two categories: (i) polar blocks, (ii) non-polar blocks. The first ones correspond, here, to the Ti-rich blocks with $x$=1, 0.875 and 0.75, and are the blocks taken into account in the simulation. The others are considered as forming a dielectric matrix, with a dielectric constant taken to be 220~\cite{bzt}. This dielectric constant is thus used to screen the dipole-dipole interactions between the polar blocks.

\begin{figure}[htbp]
    {\par\centering
    {\scalebox{0.45}{\includegraphics{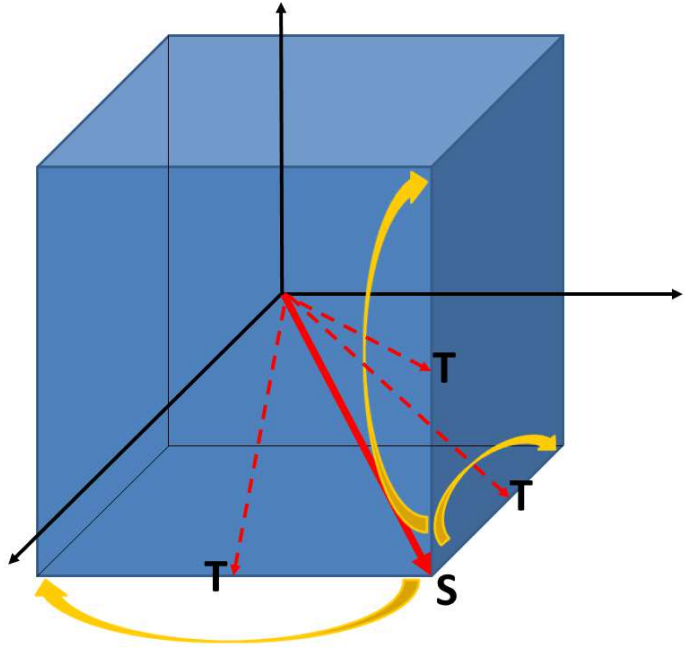}}}
    \par}
     \caption{{\small The eight possible states of ${\bf  u_k^{(N)} }$ of polar block $k$ (corners of the cube).
		Starting from one of these stable sites (S), the block variable can jump onto one of the three nearest sites, by overcoming a saddle point position located along a $<$110$>$-type direction (T site). "S" and "T" stand for "Stable" and "Transition".
     }}
    \label{hopping}
\end{figure}

In each polar block characterized by a local composition $x$, we assume 8 minima for the block variables along the $<$ 111 $>$ type directions. The block variable can thus have one among 8 possible values. At any step of the simulation, each polar block has thus its block variable in one the 8 possible states. It can thus undergo three possible hopping motions onto the three nearest sites, by overcoming a free energy barrier, with a transition state that we assume as being at the minimum of $H_{loc}^{(N)}$ along the $<$ 110 $>$-type directions. This is illustrated on Fig.~\ref{hopping}. The free energy barrier is computed as explained in the main article.

\subsubsection{Length of the KMC trajectories}
In a KMC simulation, one fixes the number of steps $N_{step}$. For a given number of steps, the length of the KMC trajectory $t$ depends on the temperature. If $\tau$ is a typical hopping time ($\propto e^{+ E_a (T) / k_B T}$, $E_a$ typical energy barrier), and $N_{cell}$ the number of polar blocks considered in the simulation ($\propto$ number of events at each step), the typical time increment is $\sim$ $\frac{\tau }{N_{cell} }$. The length of the trajectory is thus 

\begin{equation}
\label{length}
t \sim \frac{N_{step} \tau_0 e^{+ E_a  (T) / k_B T}}{N_{cell} }
\end{equation}

In the present case, we want to produce trajectories containing at least a few periods of the electric field, in order to achieve the fit of the dielectric response.

At low temperature, the events are very infrequent and the hopping times very long. Thus a few tens of thousands steps are usually sufficient to obtain a trajectory containing a large number of periods of the external field (we perform typically 50 000 steps). 
The length of the trajectory exponentially decreases with temperature. Simulations are usually possible up to the temperature $T_{max}$ corresponding to the maximum of $\chi'$, for which several million steps have to be performed. Above $T_{max}$, the elementary events are so fast that simulating the dynamics of the system over several field periods becomes impossible. However, in that case, we have made the reasonable approximation that the external field and the dielectric response evolve perfectly in phase, leading to $\chi''$=0 (this is the case of the highest temperature points in Fig.2 of the paper). Long trajectories of several million steps are nevertheless necessary even in this case to obtain $\chi'$.

\subsubsection{Dielectric response}
As explained in the main article, we reproduce the dielectric response by applying a sinusoidal external electric field ${\bf{E_{ext}}}(t) = E_0 {\bf{e_{x}}} cos(\omega t)$, ${\bf{e_{x}}}$ being the unit vector along the pseudo-cubic [100] axis. It superimposes to the internal field (related to the interactions with neighboring blocks) to change locally the free energy landscape for each block variable, and thus the hopping rates, according to $\Delta F_{loc}(T;x) \rightarrow  \Delta F_{loc}(T;x) + Z_{b}^{*} ({\bf{u_{k}}} - {\bf{u_{k,k'}}}). {\bf{E_{loc}}}(n)$, $Z_b^{*}$ being the effective charge of the block, ${\bf{u_{k}}}$ the position of stable site $k$ and ${\bf{u_{k,k'}}}$ that of the transition state from $k$ to $k'$.

Each block variable has thus the tendency to align along the external field, producing along ${\bf{e_x}}$ a macroscopic polarization $P_x(t) = P_0(\omega) cos(\omega t + \phi(\omega))$, whose amplitude $P_0(\omega)$ and phase $\phi(\omega)$ are directly related to the real and imaginary part of the dielectric susceptibility $\chi(T,\omega)$ (assuming linear response).

The delay of the response depends on the temperature, because the hoppings are thermally activated. The characteristic temperature at which the block variables freeze can be estimated by comparison of the hopping time $\tau \sim \tau_0 e^{+ \Delta F_{loc}(T) / k_B T}$ with the period of the external field $t_e = 2 \pi /  \omega$ (we assume a $\Delta F_{loc}$ not dependent on $x$ for this qualitative discussion):

\begin{itemize}

\item Low temperature: $\tau >> t_e$, the dielectric response is very low because the block variables are frozen and have not the time to follow the external solicitation. As T increases, a response appears, but delayed with respect to the field. The phase is smaller and smaller as T increases.

\item $\tau \sim t_e$: the block variables now follow the external field, there is a resonance roughly corresponding to the maximum of $\chi'$. The resonance condition roughly provides the low of evolution of $T_{max}$ with respect to $\omega$.

\item $\tau << t_e$: the block variables follow instantaneously the external field. The thermal agitation at high T tends to equalize the probabilities of the different directions of the block variables, explaining the decrease of $\chi$ with T.

\end{itemize}

The relation between $T_{max}$ and $\omega = 2 \pi f$ is driven by the resonance condition $f \sim f_0 e^{- \Delta F_{loc}(T_{max}) / k_B T_{max}}$ ($f_0 = 1/ \tau_0$), and thus mainly controlled by the local part of $H^{(N)}$, namely $H_{loc}^{(N)}$.

Two remarks are necessary:

First, the relation $f(T_{max})$ we obtain is the direct consequence of the phenomenological form we have chosen for $H_{loc}^{(N)}$, and more precisely of the temperature dependence of the hopping free energy barrier $\Delta F_{loc}(T)$. We recall that we have taken a very simple Landau-like form for $H_{loc}^{(N)}({\bf u};x)$, as a 4$^{th}$-order polynom in ${\bf u}$ and a quadratic coefficient varying linearly with $T$.

Second, in the present case, there are several hopping times depending on the local chemical composition $x$; it is thus not possible to predict analytically the relation $f(T_{max})$.

Eq.~\ref{length} shows that, for a given number of KMC steps, the length of the trajectory strongly decreases with increasing temperature. Since we are interested in performing a fit of the dielectric response, it is thus necessary to increase the number of KMC steps as T increases, in order to have a trajectory sufficiently long so as to extract $P_0$ and $\phi$ with a correct numerical precision. However, since the energy barrier strongly decreases with T, several million steps are necessary when we approach $T_{max}$ from below, and it becomes quasi-untractable to simulate at least one period of the external field above $T_{max}$. Thus, for 
the highest-temperature points of each frequency, we make the assumption that ${\bf P}$ and ${\bf  E_{ext} }$ evolve in phase, so that $\phi$ in enforced to zero in the fit. This provides $\chi''$=0 (this is the case for the high-temperature points of Fig.2c-d).

To extract $T_{max}$, the temperature at which $\chi'$ is maximum, we fit the curve $\chi'(T)$ by using Eq. (6a) of Ref.~\onlinecite{cheng1998}. Then we plot $ln f$ as a function of $T_{max}$, and fit it using an Arrhenius law, $ln f = ln f_0 - \frac{U}{k_B T_{max}}$, and using a Vogel-F\"ulcher law $ln f = ln f_0 - \frac{U}{k_B(T_{max}-T_f)}$. The values obtained for the different parameters are given below (Tab.~\ref{parameters}).

\begin{table}[h]
\small
  \caption{\ Parameters obtained by fitting the KMC data (T$_{max}$) on Arrhenius and Vogel-F\"ulcher laws, and on a form assuming a free energy barrier having a quadratic form in $T$.}
  \label{parameters}
  \begin{tabular*}{0.5\textwidth}{@{\extracolsep{\fill}}lcc|cc}
Parameter  & Arrhenius  &  Vogel-F\"ulcher & & quadratic barrier   \\
\hline
f$_0$ (Hz)  & 6.59 $\times$ 10$^{31}$    & 1.02 $\times$ 10$^{10}$  & f$_0$ (Hz) & 2.5 $\times$ 10$^{7}$ \\
U (eV)   &  0.782   &  0.019  & $\alpha$ (K$^{-1}$) & 1.89753 \\
T$_f$ (K)   &  --   &  127.602 & T$_1$ (K) &  167.333 \\	
  \end{tabular*}
\end{table}

We have to keep in mind that the relation between $f$ and $T_{max}$ reflects the phenomenological form chosen for the local thermodynamic potential $H_{loc}^{(N)}$, and that the fit is performed over a series of 4 values only. However, we obtain a fit compatible with the Vogel-F\"ulcher law, with, interestingly, a freezing temperature $T_f$ (=127.602 K) in agreement with the one determined by MC simulations using the microscopic Hamiltonian~\cite{bzt} (130 K). 

Since the local free energy barriers evolves as $\alpha(T-T_0)^2$ within our phenomenological form (see Fig. 2a of the main article), we also perform a fit by a law of the form $ln f = ln f_0 - \frac{\alpha(T_{max}-T_1)^2}{T_{max}}$ (Tab.~\ref{parameters}).

\end{document}